\documentclass[a4paper,dvipsnames,11pt]{article}
\RequirePackage{pdf14}
\RequirePackage{fix-cm}
\usepackage{./aux/jheppub}

\usepackage{graphicx}
\usepackage{verbatim}
\usepackage{amsmath,amssymb,amsfonts,amsthm,amscd}
\usepackage{mathtools}
\usepackage{bm} 
\usepackage{bbm} 
\usepackage{mathrsfs} 
\usepackage{lscape} 
\usepackage{subcaption} 
\usepackage{cancel}
\usepackage{enumitem}
\usepackage{tcolorbox}
\usepackage{multirow} 
\usepackage{colortbl} 
\usepackage{array,hhline,arydshln,multirow}
\usepackage{floatrow}
\usepackage[export]{adjustbox}
\usepackage{rotating} 
\usepackage[parfill]{parskip}
\allowdisplaybreaks[2]  
\usepackage{pifont,dsfont}
\usepackage{array,setspace,mathrsfs,yfonts,dsfont,amscd,euscript}
\usepackage{relsize,suffix,mathtools,cancel,bm,bbm}

\usepackage{graphicx,tikz}
\usepackage[framemethod=TikZ]{mdframed} 
\usepackage{tikz-cd} 
\usepackage{tikz}
\usetikzlibrary{positioning}

\usepackage{xcolor}

\definecolor{amaranth}{rgb}{0.9, 0.17, 0.31}
\definecolor{coolblack}{rgb}{0.0, 0.18, 0.39}
\definecolor{gold(web)(golden)}{rgb}{1.0, 0.84, 0.0}
\definecolor{deepcarmine}{rgb}{0.66, 0.13, 0.24}

\newcommand{\Mc}{M(0,0;c)}

\newcommand{\mN}{\mathcal{N}}
\newcommand{\mO}{\mathcal{O}}

\newcommand{\mfR}{\mathfrak{R}}

\newtheorem*{remark}{Remark}

\title{Towards a classification of graded unitary $\texorpdfstring{\mathcal{W}_3}{W3}$ algebras}

\author{Christopher Beem and}
\author{Harshal Kulkarni}

\affiliation{Mathematical Institute, University of Oxford, Woodstock Road, Oxford, OX2 6GG, UK}
\abstract{We study constraints imposed by four-dimensional unitarity (formalised as graded unitarity in recent work by the first author) on possible ${\mathcal W}_3$ vertex algebras arising from four-dimensions via the SCFT/VOA correspondence. Under the assumption that the $\mathfrak{R}$-filtration is a weight-based filtration with respect to the usual strong generators of the vertex algebra, we demonstrate that all values of the central charge other than those of the $(3,q+4)$ minimal models are incompatible with four-dimensional unitarity. These algebras are precisely the ones that are realised by performing principal Drinfel'd--Sokolov reduction to boundary-admissible $\mathfrak{sl}_3$ affine current algebras; those affine algebras were singled out by a similar graded unitarity analysis in \cite{ArabiArdehali:2025fad}. Furthermore, these particular vertex algebras are known to be associated with the $(A_2,A_q)$ Argyres--Douglas theories.}

\begin{document}
\maketitle
\flushbottom


\section{\label{sec:intro}Introduction}

Via the SCFT/VOA correspondence \cite{Beem:2013sza}, four-dimensional $\mN=2$ superconformal field theories (SCFTs) give rise to associated vertex operator algebras. The vertex algebras that arise in this fashion are known to enjoy a number of nice properties; for example, they are conjecturally quasi-Lisse so their vacuum characters satisfy finite-order linear modular differential equations \cite{Beem:2017ooy,Arakawa:2016hkg} and in many cases they admit extremely nice free field realisations \cite{Beem:2019tfp,Beem:2019snk,Beem:2021jnm,Beem:2024fom}. A particularly striking feature is that when the four-dimensional theory in question is unitary (an assumption that underlies much of the analysis of the SCFT/VOA correspondence), the associated vertex algebra inherits a novel form of unitarity dubbed \emph{graded unitarity} in \cite{ArabiArdehali:2025fad}. This structure and some of its variations were elaborated in \cite{Beem:2025guj}, where a more mathematically precise presentation can be found. (Some preliminary consequences of four-dimensional unitarity for the associated vertex operator algebra, which retrospectively are consequences of graded unitarity, were studied in \cite{Liendo:2015ofa,Beem:2018duj}).

A graded unitarity structure for a vertex algebra $\mathbb{V}$ amounts, in particular, to a conjugation operation $\rho$ (generally of order four) together with a nondegenerate half-integral good filtration $\mathfrak{R}_{\bullet}$ on $\mathbb{V}$. With respect to these ingredients, graded unitarity amounts to the requirement that a suitably defined Hermitian form is positive definite; equivalently, that the constants appearing in the two-point functions of operators with their conjugates have definite (and $R$-charge-dependent) signs.

It is an important problem to characterise to what extent the existence of a graded unitary structure constrains the underlying vertex algebra. A direct attack on the problem (for a fixed underlying vertex algebra) is complicated by the dependence of the positivity property on the choice of $\mathfrak{R}$-filtration. At present, there is no intrinsic characterisation of the physically relevant $\mathfrak{R}$-filtration without reference to four-dimensional input. One is thus led to an interesting (but seemingly difficult) problem of constraining a choice of good filtrations subject to highly specific sign requirements for two-point functions.

If, however, an $\mathfrak{R}$-filtration is given or assumed, then graded unitarity predicts a specific constellation of positivity/negativity conditions for two-point functions in the underlying vertex algebra, and these are comparatively accessible. This approach was pursued in \cite{ArabiArdehali:2025fad} for the cases where the underlying vertex operator algebra is a Virasoro VOA or an affine Kac--Moody VOA of type $\mathfrak{sl}_n$ with $n=2,3,4$. There, a simple but physically motivated choice of $\mfR$-filtration was assumed and consequently all but a discrete set of values for the central charge/level were ruled out by imposing the relevant positivity bounds level by level. In fact, the full set of positivity constraints was not used in that analysis, but only the combined positivity constraints visible at the level of the Kac determinant at a given level (and $\mathfrak{g}$-weight in the Kac--Moody case).

The purpose of the present paper is to extend this type of analysis to the case where the underlying vertex algebra is the $\mathcal{W}_3$ algebra \cite{Zamolodchikov:1985wn}. It is generally understood that the $(A_2,A_{3+3n})$ and $(A_2,A_{4+3n})$ Argyres--Douglas SCFTs give rise to the $\mathcal{W}_3$ algebra at central charges \cite{BeemCrete,RastelliHarvard,Cordova:2015nma}
\begin{equation}
    c_{3,7+3n} = 2\left(1-12\frac{(4+3n)^2}{21+9n}\right)~,\qquad c_{3,8+3n} = 2\left(1-12\frac{(5+3n)^2}{24+9n}\right)~,
\end{equation}
respectively, so these $\mathcal{W}_3$ algebras are expected to be graded unitary. These are precisely the \emph{boundary} minimal model levels, which arise by principal quantum Drinfel'd--Sokolov reduction from (boundary admissible) $\mathfrak{sl}_3$ current algebras at level $k=-3+\frac{3}{3n+4}$ and $k=-3+\frac{3}{3n+5}$. In particular, these are exactly the $\mathfrak{sl}_3$ current algebras identified in \cite{ArabiArdehali:2025fad} as the only ones not immediately inconsistent with graded unitarity (subject to an $\mfR$-filtration assumption).

At a technical level, the $\mathcal{W}_3$ case is more involved than the Virasoro and affine examples treated in \cite{ArabiArdehali:2025fad}, owing to the apparent absence of a closed-form expression in the literature for the vacuum Kac determinant of the $\mathcal{W}_3$ algebra in the form needed for a systematic sign analysis. A first contribution of this paper is therefore to derive such an expression at generic central charge. Concretely, in Section \ref{sec:W3_intro} we review the submodule structure of the vacuum Verma module $M(0,0;c)$ and the Verma resolution of the vacuum module $V(0,0;c)$, and use a Jantzen filtration argument to obtain a determinant formula for the vacuum module as a limit of an explicit ratio of products of Verma determinants together with certain $c$-dependent normalisation factors \eqref{vacdet}. We then simplify this expression to a manifestly rational form in terms of linear factors $(c-c_{p,q})$ with explicitly computable exponents \eqref{rationalc}.

With this closed-form determinant formula in hand, we return to graded unitarity. As in \cite{ArabiArdehali:2025fad}, we take as our diagnostic the determinant-level consequence of positivity: at each conformal level $N$, graded unitarity predicts a definite sign for the determinant of the Shapovalov form, determined by the distribution of $R$-charges (after refining the $\mfR$-filtration to a grading). Since deriving the $R$-filtration from first principles is currently out of reach, we make the plausible assumption that the $R$-filtration is weight-based with respect to the strong generators $T$ and $W$, in the sense made precise in Section \ref{sec:preliminary_constraints}. Under this assumption, we can determine the graded-unitarity prediction for the sign of the level-$N$ vacuum determinant for each level and compare it with the sign implied by the factorised expression \eqref{rationalc}.

Our main result is that this comparison is extremely restrictive: under the above $\mfR$-filtration hypothesis, all values of the central charge other than those of the $(3,q+4)$ minimal models are incompatible with four-dimensional unitarity. Thus it is only the $\mathcal{W}_3$ algebras arising from the $(A_2,A_q)$ Argyres--Douglas theories that are apparently allowed.

More broadly, the present results add to a growing body of evidence that four-dimensional unitarity acts as a powerful rigidity principle for the vertex algebras appearing in the image of the SCFT/VOA map. In each class of examples studied thus far---Virasoro, affine $\mathfrak{sl}_{2,3,4}$, and now $\mathcal{W}_3$---even the determinant-level consequences of graded unitarity already single out (or nearly single out) the boundary or minimal model values of central charges that are known to arise from four-dimensional theories. At the same time, the determinant test used here captures only a comparatively coarse shadow of the full positivity constraints encoded by graded unitarity. Clearly, a more conceptual understanding of graded unitarity within vertex algebra theory is needed. In particular, it would be desirable to identify the intrinsic structural features of a vertex algebra that give rise to (or obstruct) the existence of a compatible $\mfR$-filtration and conjugation satisfying the relevant positivity conditions.

Our derivation of a closed-form vacuum determinant for the ${\mathcal W}_3$ algebra seems to suggest that analogous formulae for higher ${\mathcal W}_N$ algebras may be obtainable. Although the combinatorics of singular vectors and the associated Kazhdan--Lusztig data become increasingly intricate for $N>3$, the Jantzen methods employed here appear sufficiently robust to make such generalisations feasible. We expect that the resulting determinant formulae will provide similarly strong restrictions on ${\mathcal W}_N$ central charges when assuming graded unitarity and plausible $\mfR$-filtrations \cite{WNtoappear}.

Finally, taken together with the $\mathfrak{sl}_3$ analysis of \cite{ArabiArdehali:2025fad}, the present results raise a natural structural question: how does a graded unitary structure behave under quantum Drinfel'd--Sokolov reduction? Though there are some naive obstacles (from both an SCFT and a vertex algebra perspective) to following this structure through reduction, we are hopeful that a careful treatment (perhaps subject to some technical niceness assumptions) might allow a graded unitarity structure to be transported to the reduced vertex algebra.

The organisation of the paper is as follows. In Section \ref{sec:W3_intro} we review the relevant structure of the $\mathcal{W}_3$ algebra and its Verma modules, determine the singular-vector/submodule structure of the vacuum Verma module, and derive our vacuum determinant formula. In Section \ref{sec:preliminary_constraints} we specialise the general framework of graded unitarity to the $\mathcal{W}_3$ setting and describe the weight-based $\mfR$-filtrations we will consider. In Section \ref{sec:classification} we formulate the determinant-level sign constraint and use it to test graded unitarity: we begin with low-level computations as a check on the general determinant formula, analyse the large negative-$c$ regime to confirm compatibility with the predicted sign, and then study the structure of zeros of the vacuum Kac determinant to deduce the central charge constraints culminating the aforementioned restriction. Appendices \ref{app} and \ref{app2} contain computational details and the simplification of the Kac determinant expression.

\section{\label{sec:W3_intro}Relevant aspects of the \texorpdfstring{${\mathcal W}_3$}{W3} algebra}

We are exclusively interested in the ${\mathcal W}_3$ algebra \cite{Zamolodchikov:1985wn}. For generic central charge $c$, this is the unique vertex operator algebra strongly generated by bosonic currents $\{T,W\}$ of conformal weight two and three, respectively. The singular OPEs of these strong generators take the following well-known form,
\begin{equation}\label{eq:W3_OPE}
\begin{split}
    T(z)T(w)&\sim \frac{\frac{c}{2}}{(z-w)^4}+\frac{2T(w)}{(z-w)^2}+\frac{\partial T(w)}{z-w}~,\\
    T(z)W(w)&\sim \frac{3W(w)}{(z-w)^2}+\frac{\partial W(w)}{z-w}~,\\
    W(z)W(w)&\sim \frac{\frac{c}{3}}{(z-w)^6}+\frac{2T(w)}{(z-w)^4}+\frac{\partial T(w)}{(z-w)^3}\\
    &\qquad\qquad\qquad+\frac{\frac{32}{22+5c}\Lambda(w)+\frac{3}{10}\partial^2T(w)}{(z-w)^2}+\frac{\frac{16}{22+5c}\partial\Lambda(w)+\frac{1}{15}\partial^3T(w)}{z-w}~,
\end{split}
\end{equation}
where $\Lambda(w)\coloneqq(TT)(w)-\frac{3}{10}\partial^2T(w)$ is the weight-four quasi-primary composite of two stress tensors. Upon expanding the generating fields into Fourier modes,
\begin{equation}
    T(z)=\sum_{m\in\mathbb{Z}}L_{-m-2}z^m~,\qquad W(z)=\sum_{m\in\mathbb{Z}}W_{-m-3}z^m~,
\end{equation}
the singular OPEs \eqref{eq:W3_OPE} translate into the following commutation relations,
\begin{equation}
\begin{split}
    [L_n, L_m] &= (n-m) L_{n+m} + \frac{c}{12} n (n^2 - 1) \delta_{n+m,0}~,\\ 
    [L_n, W_m] &= (2 n - m) W_{n+m}~,\\
    [W_n, W_m] &= \frac{c}{360} (n^2 - 4) (n^2 - 1) n \delta_{n+m, 0} + \frac{16}{22 + 5 c} (n-m) \Lambda_{n+m}\\ 
    &\qquad+ (n-m) \left[ \frac{1}{15} (n+m+2)(n+m+3) - \frac{1}{6} (n+2)(m+2) \right] L_{n+m}~,
\end{split}
\end{equation}
where $\Lambda_n$ denote the modes of the quasi-primary $\Lambda(z)$,
\begin{align}
    \Lambda_n = \sum_{k = -\infty}^\infty : L_k L_{n-k}: + \frac{1}{5} x_n L_n
\end{align}
with
\begin{equation}
    x_{2l} = (1+l)(1-l)~,\qquad x_{2 l+1} = (2+l)(1-l)~,\qquad l \in \mathbb{Z}~.
\end{equation}

For generic $c$, this vertex algebra is freely generated by $T$ and $W$ (in the sense of \cite{sole2005freely}), whereas for central charges given by
\begin{equation}
    c=c_{p,q} = 2\left(1-12\frac{(p-q)^2}{pq}\right)~,\qquad p,q\in\mathbb{Z}_{>0}~,\qquad (p,q)=1~,
\end{equation}
the universal (freely generated) ${\mathcal W}_3$ algebra becomes reducible and there exist nontrivial quotients. In this paper, we will be interested in both generic and non-generic $c$, but for our analysis it will always be sufficient to consider the universal/generic ${\mathcal W}_3$ algebra.

\subsection{Verma modules and the vacuum module}

Denote by $|h,w\rangle$ a highest-weight state obeying the highest-weight conditions
\begin{equation}
\begin{split}
    L_n |h, w\rangle &= W_n|h,w\rangle = 0~,\qquad n>0~, \\
    L_0|h,w\rangle &= h |h,w\rangle~,\\
    W_0 |h,w\rangle &= w |h, w\rangle~.
\end{split}
\end{equation}
We will call a highest weight state with $h = w = 0$ a \emph{vacuum} and denote it by $|0\rangle$. 

We denote by $M(h,w;c)$ the ${\mathcal W}_3$ Verma module with highest weights $h$ and $w$. This is a module for the ${\mathcal W}_3$ algebra at central charge $c$ with highest weight state $|h,w\rangle$; it can be equipped with a PBW-style basis consisting of states of the form
\begin{equation}\label{eq:PBW_basis_W3}
    \{L_{-i_1} L_{-i_2} \ldots L_{-i_n} W_{-j_1} W_{-j_2} \ldots W_{-j_m}|h,w\rangle \}~,
\end{equation}
with $i_1 \geqslant i_2 \geqslant \ldots \geqslant i_n > 0$ and $j_1 \geqslant j_2 \geqslant \ldots \geqslant j_m > 0$.\footnote{There is nothing privileged about this choice of PBW ordering, and indeed in other references (\emph{cf.} \cite{Mizoguchi:1988vk,Carpi:2019szo}) different monomial basis elements (in the negatively-moded generators) are used. This will not be consequential for anything that follows.} For generic $c,h,w$ the Verma module $M(h,w;c)$ is simple.

Verma modules admit a unique-up-to-rescaling invariant Hermitian form (sometimes called the Shapovalov form), where invariance amounts to the rule that the Hermitian conjugates of modes $L_n$ and $W_n$ are given by
\begin{equation}
    L_n^\dag = L_{-n}~,\qquad W_n^\dag = W_{-n}~.
\end{equation}
We will denote by $\langle-|-\rangle$ the Shapovalov form normalised so that $\langle h,w|h,w\rangle=1$ for the specified highest-weight vector.

The Kac determinant for ${\mathcal W}_3$ Verma modules (\emph{i.e.}, the determinant of the Shapovalov form, evaluated in a monomial basis such as \eqref{eq:PBW_basis_W3}) has been known since the 1980's \cite{Mizoguchi:1988vk, Bouwknegt:1992wg, Watts:1989bn}. Denoting by $G^{(N)}(h,w;c)$ the Gram matrix at level $N$, the resulting determinant is given by\footnote{This expression is independent of the choice of PBW monomial basis--indeed for any such monomial basis there is an ordering of the basis elements such that a change of PBW monomials is unitriangular, so the determinant of the corresponding Gram matrix is unaffected.}
\begin{equation} \label{Vermadet}
    \det G^{(N)}(h,w;c) = A_N \prod_{k = 1}^N \prod_{\substack{n,m\in\mathbb{N}\\nm= k}} \left( f_{n,m}(h,c) - w^2 \right)^{P_2(N-k)}~,
\end{equation}
where $A_N \in \mathbb{R}_{\geqslant 0}$ is independent of $h$, $w$, and $c$ (and depends on an overall real, positive choice of normalisation for the highest weight state), and $P_2(N)$ counts the number of states at level $N$,
\begin{equation}
    P_2(N) \coloneqq \sum_{k = 0}^N p(k) p(N-k)~,
\end{equation}
where $p(i)$ is the classical partition function. The expression for $f_{n,m}(h,c)$ takes the form
\begin{equation} \label{eqnfmn}
\begin{split}
    f_{n,m}(h,c) = \frac{64}{9(22 + 5 c)} &\left( h + (4-n^2) \alpha_+^2 + (4-m^2) \alpha_-^2 - 2 + \frac{1}{2}nm \right) \\
    &\times \left(h - 4\left[(n^2-1)\alpha_+^2 + (m^2-1)\alpha_-^2 \right] - 2(1 - nm)\right)^2~,
\end{split}
\end{equation}
where
\begin{equation}
    \alpha_\pm^2 = \frac{50 - c \pm \sqrt{(2-c)(98-c)}}{192}~.
\end{equation}

The ${\mathcal W}_3$ vacuum module $V\coloneqq L(0,0;c)$ at generic $c$ is the quotient of the Verma module $M(0,0;c)$ by its (maximal) submodule which is generated by $L_{-1}|0\rangle$ and $W_{-1}|0\rangle$. The highest weight state of $L(0,0;c)$ thus additionally obeys $L_{-1}|0\rangle = W_{-1}|0\rangle=W_{-2}|0\rangle = 0$, and $L(0,0;c)$ admits a PBW-style basis consisting of states of the form
\begin{equation}
    \{L_{-i_1} L_{-i_2} \ldots L_{-i_n} W_{-j_1} W_{-j_2} \ldots W_{-j_m}|0,0\rangle \}
\end{equation}
with $i_1 \geqslant i_2 \geqslant \ldots \geqslant i_n \geqslant 2$ and $j_1 \geqslant j_2 \geqslant \ldots \geqslant j_m \geqslant 3$. As far as the authors have been able to determine, an explicit expression for the Kac determinant of the ${\mathcal W}_3$ vacuum module has not appeared in the literature. Deriving such an expression is the purpose of the remainder of this section.

\subsection{\label{subsec:submodules_and_singulars}Submodules and singular vectors of \texorpdfstring{$M(0,0;c)$}{M(0,0;c)}}

To derive a determinant formula for the vacuum module, it will be helpful to understand the structure of ($i$) singular vectors and submodules for the Verma module $M(0,0;c)$ of which the vacuum module is a quotient, and ($ii$) the resolution of the vacuum module itself by Verma modules. At generic central charge, both of these structures are controlled by analogous relations for $\mathfrak{sl}_3$-modules (since this controls the analogous structures for affine $V^k(\mathfrak{sl}_3)$ modules at generic level, and these in turn imply the ${\mathcal W}_3$ case by applying the Drinfel'd--Sokolov reduction functor\footnote{The authors thank Gurbir Dhillon for explanations on this point.}). Nevertheless, we will give a detailed account directly in the ${\mathcal W}_3$ setting, which will give us additional perspective on certain important but idiosyncratic normalisation factors.

It is also convenient to introduce a different basis for the singular level-one subspace of $M(0,0;c)$, for which $W_0$ is also diagonalised. Noting the $W_0$ action,
\begin{equation}
\begin{split}
    W_0 L_{-1} |0,0\rangle &= 2 W_{-1} |0,0\rangle~,\\
    W_0 W_{-1} |0,0\rangle &= \frac{(2-c)}{22+5c} L_{-1} |0,0\rangle~,
\end{split}
\end{equation}
one can form $W_0$ eigenstates,
\begin{equation}
    |\chi_\pm\rangle \equiv |1,w_\pm\rangle \coloneqq L_{-1}|0,0\rangle \pm \sqrt{\frac{2(22 + 5c)}{2 - c}} W_{-1}|0,0\rangle~,
\end{equation}
where $w_\pm$ are the $W_0$ weights of $|\chi_\pm\rangle$ given by
\begin{equation}
    w_\pm = \pm \sqrt{\frac{2(2-c)}{22+5c}}~.
\end{equation}
This diagonalisation requires $c \neq 2$. Ultimately, (four-dimensional) unitarity will restrict the central charge of ${\mathcal W}_3$ to satisfy $c = 0$ or $c \leqslant -\frac{22}{5}$, so any pathologies at $c=2$ will not be relevant for us here.

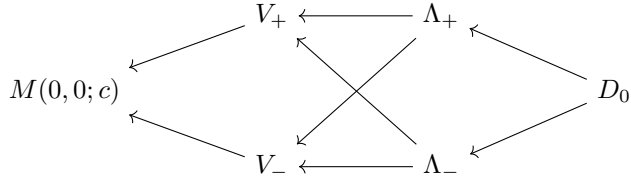
\begin{figure}
\centering
\begin{tikzpicture}[
  node distance=20mm and 15mm,
  every node/.style={font=\small}
]

\node (M) { $M(0,0;c)$ };

\node[right=of M, yshift=10mm] (Vplus) { $V_+$ };
\node[right=of M, yshift=-10mm] (Vminus) { $V_-$ };

\node[right=of Vplus] (Wplus) { $\Lambda_+$ };
\node[right=of Vminus] (Wminus) { $\Lambda_-$ };

\node[right=of M, xshift=45mm] (D) { $D_{0}$ };

\draw[->] (Vplus) -- (M);
\draw[->] (Vminus) -- (M);

\draw[->] (Wplus) -- (Vplus);
\draw[->] (Wplus) -- (Vminus);

\draw[->] (Wminus) -- (Vplus);
\draw[->] (Wminus) -- (Vminus);

\draw[->] (D) -- (Wplus);
\draw[->] (D) -- (Wminus);

\end{tikzpicture}

\caption{\label{fig:submod-structure}
Diagram representing the embedding of singular vectors/submodules in the vacuum Verma module $\Mc$ at generic central charge. Arrows indicate the embedding of a singular vector (the highest weight vector of the source submodule) into the target submodule.}
\end{figure}

Each of $|\chi_\pm\rangle$ is the highest weight state of a submodule of $M(0,0;c)$, which we denote by $V_\pm$. The corresponding Verma modules $M(1,w_\pm;c)$ (with the same highest weights) each have exactly two singular vectors (both at level two, so with $L_0$ weight $h=3$), and these singular vectors are present in the submodules $V_\pm$ (\emph{i.e.}, they are not identically zero), so in fact we have the identification
\begin{equation}
    V_\pm \cong M(1,w_\pm;c)~.
\end{equation}
However, the submodules $V_+$ and $V_-$ have a non-trivial intersection. Denote by 
\begin{equation}
    |\xi^{(1)}_\pm\rangle \equiv |3, \lambda_\pm^{(1)} \rangle,
\end{equation}
the two weight-three singular vectors in $V_+$, and by
\begin{equation}
    |\xi^{(2)}_\pm\rangle \equiv |3, \lambda_\pm^{(2)} \rangle,
\end{equation}
the two weight-three singular vectors in $V_-$, where the $W_0$ weights are given by
\begin{equation}
    \lambda^{(1)}_+ = \lambda^{(2)}_+ = \lambda_+\coloneqq +\sqrt{\frac{2(98-c)}{22+5c}}~,\qquad \lambda^{(1)}_- = \lambda^{(2)}_- =\lambda_-\coloneqq  -\sqrt{\frac{2(98-c)}{22+5c}}~.
\end{equation}
Then we have the following identification in $M(0,0;c)$ (up to normalisation choices),
\begin{equation}
    |\xi^{(1)}_+\rangle = |\xi^{(2)}_+\rangle\eqqcolon |\xi_+\rangle~,\qquad
    |\xi^{(1)}_-\rangle = |\xi^{(2)}_-\rangle\eqqcolon|\xi_-\rangle~.
\end{equation}
Let $\Lambda_\pm$ denote the submodule of $M(0,0;c)$ generated by $|\xi_\pm\rangle$. The corresponding Verma module $M(3,\lambda_\pm;c)$ has exactly one level-one (so $h=4$) singular vector, and this is again present in the submodule so $\Lambda_\pm$ are isomorphic to Verma modules, 
\begin{equation}
    \Lambda_{\pm}\cong M(3,\lambda_\pm;c)~,
\end{equation}
and furthermore these are identified with one another (as with their highest weight states),
\begin{equation}
    \Lambda^{(1)}_\pm = \Lambda^{(2)}_\pm\eqqcolon \Lambda_\pm~.
\end{equation}
The submodules $\Lambda_+$ and $\Lambda_-$ again have a non-trivial intersection. The Verma modules $M(3,\lambda_\pm;c)$ each have a single level-one (so $h = 4$) singular vector having $W_0$ weight zero, which we denote as $|\zeta_\pm\rangle$ and the submodules generated by these singular vectors we denote as $D_\pm$. These are simple and isomorphic to Verma modules,
\begin{equation}
    D_\pm \cong M(4,0;c)~.
\end{equation}
The level-one singular vectors inside $\Lambda_+$ and $\Lambda_-$ are identified in $M(0,0;c)$, so we have
\begin{equation}
    |\zeta_+\rangle = |\zeta_-\rangle\eqqcolon|\zeta\rangle~,
\end{equation}
which leads to the identification of the submodules,
\begin{equation}
    D_+ = D_- \eqqcolon D_0
\end{equation}
The full submodule structure for the ${\mathcal W}_3$ vacuum Verma module is represented in Figure \ref{fig:submod-structure}. As foreshadowed above, this has precisely the structure of the strong Bruhat order on the Weyl group of $\mathfrak{sl}_3$, which controls the (equivalent) structure of embedded singular vectors in the $\mathfrak{sl}_3$ Verma module with highest weight equal to zero.


\subsection{Verma resolution of the vacuum module}

Complementary to the submodule structure reviewed above, the ${\mathcal W}_3$ vacuum module admits a resolution by Verma modules. This is again controlled by the finite $\mathfrak{sl}_3$ case (in particular, the resolution of the one-dimensional $\mathfrak{sl}_3$ module by Verma modules), which is again encoded in the strong Bruhat order.

It is easy to see the structure of the Verma resolution at the level of characters. The character of the Verma module $M(h,w;c)$ of the ${\mathcal W}_3$ algebra is given by \cite{Bouwknegt:1992wg}
\begin{equation}
    \text{ch}_{M}(q) = \text{Tr}_{M} (q^{L_0 - \frac{c}{24}}) = \frac{q^{h-\frac{c}{24}}}{\prod_{k\geqslant 1} (1-q^k)^2}~.
\end{equation}
Alternatively, the character of the vacuum module $V(0,0;c)$ is given by an analogous formula where the terms in the denominator corresponding to action of $L_{-1}$, $W_{-1}$, and $W_{-2}$ are omitted,
\begin{equation}\label{charvac}
\begin{split}
    \text{ch}_V(q) = \text{Tr}_V (q^{L_0 - \frac{c}{24}}) &= \frac{q^{-\frac{c}{24}}}{\prod_{k\geqslant 2} (1-q^k)\prod_{k\geqslant 3} (1-q^k)}\\ 
    &= q^{-\frac{c}{24}} \frac{(1-q)^2(1-q^2)}{\prod_{k\geqslant 1} (1-q^k)^2}\\
    &= q^{-\frac{c}{24}} \frac{(1-2q + 2q^3 - q^4)}{\prod_{k\geqslant 1} (1-q^k)^2}\\
    &= \text{ch}_{M_0}(q) - 2 \text{ch}_{M_1}(q) + 2 \text{ch}_{M_3}(q) - \text{ch}_{M_4}(q)~.
\end{split}
\end{equation}
Indeed, given the structure of embedded singular vectors in the vacuum Verma module reviewed above, one can identify the following resolution of the vacuum module,
\begin{equation} \label{resol}
    \begin{tikzpicture}[baseline=(current bounding box.center),
        node distance=7mm and 7mm,
        every node/.style={inner sep=2pt}
    ]

    \node (Zero1) {$0$};
    \node[right=of Zero1] (M4) {$\scriptstyle{M(4,0;c)}$};
    \node[right=of M4] (M3) {$\substack{M(3,\lambda_+;c)\\ \oplus\\M(3,\lambda_-;c)}$};
    \node[right=of M3] (M2) {$\substack{M(1,w_+;c) \\ \oplus \\M(1,w_-;c)}$};
    \node[right=of M2] (M1) {$\scriptstyle{M(0,0;c)}$};;
    \node[right=of M1] (V) {$\scriptstyle{V(0,0;c)}$};
    \node[right=of V] (Zero) {$0$};

    \draw[->] (Zero1) -- (M4) node[midway, above] {$\scriptstyle{d_5}$};
    \draw[->] (M4) -- (M3) node[midway, above] {$\scriptstyle{d_4}$};
    \draw[->] (M3) -- (M2) node[midway, above] {$\scriptstyle{d_3}$};
    \draw[->] (M2) -- (M1) node[midway, above] {$\scriptstyle{d_2}$};
    \draw[->] (M1) -- (V) node[midway, above] {$\scriptstyle{d_1}$};
    \draw[->] (V) -- (Zero) node[midway, above] {$\scriptstyle{d_0}$};
    
    \end{tikzpicture}
\end{equation}

\subsection{\label{subsec:Jantzen_determinant}Jantzen filtration and a vacuum determinant formula}

To motivate a determinant formula for the vacuum module $V(0,0;c)$, there is a useful construction that is associated with the so-called \emph{Jantzen filtration} on the Verma module $M\coloneqq M(0,0;c)$. Informally speaking, one wants to consider deforming infinitesimally away from $h=0$, and then to carefully understand the behaviour of the Verma module Kac determinant in the limit $h\to0$ in order to remove the contributions of null vectors and produce a determinant formula for the vacuum module.

To do this formally, introduce the free variable $T$ and the corresponding polynomial ring $A=\mathbb{C}[T]$, and define the ``deformed'' Verma module $M_A\coloneqq M(0+T,0;c)$ as a module over the vertex algebra ${\mathcal W}_3 \otimes_\mathbb{C}A$. The $A$-valued Shapovalov form on $M_A$ is then nondegenerate (\emph{i.e.}, the deformation in the direction of the conformal weight is generic).

One defines a decreasing filtration on $M(0+T,0;c)$ by the minimum order (in $T$) of the Shapovalov form evaluated on a given vector, \emph{i.e.},
\begin{equation}
    M_A(i) = \{v\in M_A:\quad\forall w\in M_A~,~\langle v|w\rangle\in T^i A\}~.
\end{equation}
This can be thought of as an \emph{order-of-vanishing} filtration for the Shapovalov form. This further induces a filtration of $M$ obtained by setting
\begin{equation}
    M(i) = M_A(i)\,/\,T\cdot M_A(i)~,
\end{equation}
(which can be thought of as evaluating $T$ to be zero). This is the Jantzen filtration for $M$,
\begin{equation}
    M=M(0)\supseteq M(1)\supseteq M(2)\supseteq\cdots\supseteq M(n)=0~.
\end{equation}
The non-zero quotients $M(i)/M(i+1)$ are semisimple, with nondegenerate Shapovalov form induced by the form $T^{-i}\langle-|-\rangle$ on $M_A(i)$. In particular, this means that the quotient $M(i)/M(i+1)$ represents $v \in M(h)$ for which the vanishing of the Shapovalov form in the $h \to 0$ limit is precisely of order $h^i$. As a result, the order of vanishing of the Gram determinant of the Shapovalov form on $M$ in the limit $h \to 0$ is given by 
\begin{equation} \label{dimform}
    \sum_{i=1} i \ \dim\left(M(i)/M(i+1)\right) = \sum_{i=1} \dim\left(M(i)\right)~,
\end{equation}
which is a version of Jantzen's sum rule.

The Jantzen filtration for the ${\mathcal W}_3$ vacuum Verma module can be determined directly to be as follows,
\begin{equation}\label{eq:Jantzen_assignment}
\begin{split}
    M(0)&=M~,\\
    M(1)&=V_+ + V_-~,\\
    M(2)&= \Lambda_+ + \Lambda_-,\\
    M(3)&= D_0~,\\
    M(i)&=0~,\quad i\geqslant 4~.
\end{split}
\end{equation}
in other words, it is the same as the radical/socle/Loewy filtrations of $M(0,0;c)$.

The structure of the Jantzen filtration can be determined by direct calculation of the $A$-valued norms of the singular vectors appearing in \ref{fig:submod-structure}. Alternatively, it can be inferred without direct calculation (but using the generic Verma module Kac determinant) as follows. The order of vanishing of the Verma module Kac determinant as $h\to0$ is $2P_2(N-1) + P_2(N-4)$, so by equation \eqref{dimform}, we must have
\begin{equation} \label{ordVan}
    2P_2(N-1) + P_2(N-4) = \sum_{i=1} \dim M(i)~.
\end{equation}
This identity holds precisely for the assignment given in \eqref{eq:Jantzen_assignment}. Moreover, \eqref{eq:Jantzen_assignment} is the \emph{minimal} filtration assignment compatible with the structure of embedded singular vectors, and any other assignment would involve placing composition factors deeper into the filtration and thus increase the left hand side of \eqref{dimform}. Thus the given assignment is the unique one compatible with the structure of singular vector embeddings.

Now the level $N$ Gram matrix of the Shapovalov form on $M_A$ can (by a suitable unitriangular change of basis, so the determinant will be unchanged) be adapted to the order-of-vanishing filtration, after which it will have the following block-filtered form,

\begin{equation}
G^{(N)}(T,0;c) =
\left(
\begin{tikzpicture}[baseline=(current bounding box.center)]
  \def\blocksize{1.5cm}

  \foreach \x in {1,2,3} {
    \draw[dotted] (\x*\blocksize,0) -- (\x*\blocksize,4*\blocksize);
  }
  \foreach \y in {1,2,3} {
    \draw[dotted] (0,\y*\blocksize) -- (4*\blocksize,\y*\blocksize);
  }

  \node at (0.5*\blocksize,3.5*\blocksize) {$G_0^{(N)}$};
  \node at (1.5*\blocksize,3.5*\blocksize) {$\mO(T)$};
  \node at (2.5*\blocksize,3.5*\blocksize) {$\mO$$(T^2)$};
  \node at (3.5*\blocksize,3.5*\blocksize) {$\mO$$(T^3)$};

  \node at (0.5*\blocksize,2.5*\blocksize) {$\mO(T)$};
  \node at (1.5*\blocksize,2.5*\blocksize) {$G_1^{(N)}$};
  \node at (2.5*\blocksize,2.5*\blocksize) {$\mO(T^2)$};
  \node at (3.5*\blocksize,2.5*\blocksize) {$\mO(T^3)$};

  \node at (0.5*\blocksize,1.5*\blocksize) {$\mO(T^2)$};
  \node at (1.5*\blocksize,1.5*\blocksize) {$\mO(T^2)$};
  \node at (2.5*\blocksize,1.5*\blocksize) {$G_2^{(N)}$};
  \node at (3.5*\blocksize,1.5*\blocksize) {$\mO(T^3)$};

  \node at (0.5*\blocksize,0.5*\blocksize) {$\mO(T^3)$};
  \node at (1.5*\blocksize,0.5*\blocksize) {$\mO(T^3)$};
  \node at (2.5*\blocksize,0.5*\blocksize) {$\mO(T^3)$};
  \node at (3.5*\blocksize,0.5*\blocksize) {$G_3^{(N)}$};

  \draw[dotted,very thick] (0,3*\blocksize) rectangle (1*\blocksize,4*\blocksize); 
  \draw[dotted,very thick] (1*\blocksize,2*\blocksize) rectangle (2*\blocksize,3*\blocksize); 
  \draw[dotted,very thick] (2*\blocksize,1*\blocksize) rectangle (3*\blocksize,2*\blocksize); 
  \draw[dotted,very thick] (3*\blocksize,0*\blocksize) rectangle (4*\blocksize,1*\blocksize); 
\end{tikzpicture}
\right)~.
\end{equation}

The on-diagonal blocks $G^{(N)}_i$ are Gram matrices for the ($A$-valued) Shapovalov form evaluated on (a section of) the quotient $M_A(i)/M_A(i+1)$ at level $N$. As above, $G_i^{(N)}/T^i$ becomes a non-degenerate form when evaluating $T$ to zero, and this is identified (\emph{up to $c$-dependent normalisation}) with the corresponding Gram matrix for the ($\mathbb{C}$-valued) Shapovalov form on the simple module $M(i)/M(i+1)$. In particular, the Gram matrix $G_0^{(N)}$ becomes precisely the Gram matrix for the vacuum module at level $N$. This allows us to express the vacuum Kac determinant in terms of the determinant for the vacuum Verma module as well as the determinant of nondegenerate forms for the (semi-)simple modules appearing as composition factors of the vacuum Verma,
\begin{equation} \label{jantzendet}
    \det(G^{(N)}_0(c)) = \lim_{T\to 0} \frac{\det\,G^{(N)}(T,0;c)}{\prod_{i = 1}^3 \det\,G_i^{(N)}}~,
\end{equation}
as the contributions of the off-diagonal blocks vanish in the limit. The limit is well defined and non-zero precisely because the leading power of $T$ cancels between the numerator and denominator owing to the Jantzen sum rule. 

To turn this into a useful expression, we iterate the above process where now we express the determinant of the subleading diagonal blocks $G_i^{(N)}$ at leading order in $T$ (\emph{i.e.}, the determinant of the Shapovalov form on the corresponding semi-simple composition factors) in terms of (non-vacuum) Verma module determinants, since we are already equipped with an explicit expression for the latter. Indeed, there is an exact sequence analogous to \eqref{resol} for the Verma resolution of each irreducible composition factor, and these ultimately allow us to express the level $N$ vacuum Kac determinant as a ratio of determinants of Verma modules plus normalisation factors. Details of this procedure are given in Appendix \ref{app}, culminating in the following expression for the vacuum Kac determinant at level $N$ (up to an overall positive constant),\footnote{Note that one could equivalently write $w_-$ (instead of $w_+$) and $\lambda_-$ (instead of $\lambda_+$) in the arguments of Verma determinants in \eqref{vacdet} as the Verma module Kac determinants agree for all such choices. }
\begin{equation}\label{vacdet}
\begin{split}
    \det V_N(c) \sim \frac{f_3(c)^{P_2(N-3)}}{f_1(c)^{P_2(N-1)} f_4(c)^{P_2(N-4)}} \lim_{h \to 0} \frac{1}{h^{\alpha}}\frac{\det G^{(N)}(h,0)\det G^{(N-3)}(h+3,\lambda_+)^2}{\det G^{(N-1)}(h+1,w_+)^2\det G^{(N-4)}(h+4,0)}~,
\end{split}
\end{equation}
where
\begin{equation}
    \alpha = 2P_2(N-1) - 4 P_2(N-3) + 3P_2(N-4)~,
\end{equation}
and $f_1(c)$, $f_3(c)$, and $f_4(c)$ correspond to certain normalisation factors which can be expressed in terms of ratios of $A$-valued Gram determinants of singular subspaces. Alternatively, these are fixed by explicitly computing low-level vacuum Kac determinants (in this case, say, computing the left hand side of \eqref{vacdet} up to level four). They are given by the following rational functions of the central charge,
\begin{equation}\label{norm}
\begin{split}
    f_1(c) &= \frac{2-c}{22+5 c}~,\\
    f_3(c) &= \frac{(c^2-124c-44)^2 (2-c)^2}{(98-c)(c+10)^2(22+5c)^3}~,\\
    f_4(c) &= \frac{(c-146)^2(c+10)}{(110-c)(22+5c)^2}~,
\end{split}
\end{equation}
It is straightforward to verify that this expression for the level $N$ vacuum Kac determinant matches the result of brute-force computation (see below in \ref{lowlevel}) for levels $N = 2, \ldots 10$.

The expression \eqref{norm} can be rewritten as a manifestly rational function of $c$,
\begin{align} \label{rationalc}
    \det V_N(c) \sim (22+5c)^{n(N)}\prod_{\substack{q>p\geqslant 2\\(p,q) = 1}} (c - c_{p,q})^{E^{(N)}_{p,q}}
\end{align}
where the exponents $E_{p,q}^{(N)}$ can be expressed in terms of two-coloured partition functions,
\begin{equation}\label{exponent}
\begin{split}
    E_{p,q}^{(N)} = \sum_{j \in \mathbb{Z}/\{0\}}^\prime &\left(P_2(N - (jp-2)(jq-2)) + 2 P_2(N-(jp-1)(jq-1)) \right.\\
    &\left. - 2P_2(N-1-(jp-2)(jq-1))  - 2P_2(N-1-(jp-1)(jq-2)) \right.\\
    &\left. -  2P_2(N-1-(jp-1)(jq+1)) + 2 P_2(N - 3 - (jp-2)(jq+1))\right.\\
    &\left. + 2 P_2(N - 3 - (jp+1)(jq-2)) + 2 P_2(N - 3 - (jp-1)(jq-1))\right.\\
    &\left. - P_2(N-4 - (jp-2)(jq+2)) - 2 P_2(N - 4 - (jp-1)(jq+1))\right)~.
\end{split}
\end{equation}
The exponent on the prefactor $(22+5c)^{n(N)}$ is given by
\begin{equation}
    n(N) = P_2(N-1) - 3 P_2(N-3)+ 2 P_2(N-4) - s(N) + 2s(N-1) - 2s(N-3) + s(N-4)~,
\end{equation}
where 
\begin{equation}\label{sdefn}
    s(i) = \sum_{k =1}^i \sum_{mn = k} P_2(N-k)~.
\end{equation}
Within the sum, terms of the form $P_2(N-k-(jp+a)(jq+b))$ (with $k=1,3,4$) should only be included when $(jp-a)(jq-b)>0$ (this is indicated by the prime on the sum in \eqref{exponent}). The details of the calculation of this simplified form of the vacuum determinant are given in Appendix \ref{app2}.

\medskip

\begin{remark}
We expect that the approach taken here to computing the vacuum module Kac determinant as a ratio of products of Verma module Kac determinants can be extended to more general ${\mathcal W}_N$ algebras. For $N \geqslant 4$, the existence of non-trivial Kazhdan-Lusztig polynomials for $\mathfrak{sl}_N$ requires some careful treatment of additional normalisation factors, though ultimately these could be determined by brute force calculation at low levels.

A familiar special case where this approach is easily implemented is when $N=2$, \emph{i.e.}, the Virasoro VOA. Here, the Verma module $M(0;c)$ has a level-one singular vector at generic $c$, and the submodule generated by this singular vector is isomorphic to the Verma module $M(1;c)$, which in turn does not have additional singular vectors for generic $c$. The vacuum module then admits a resolution by Verma modules with only three terms. In this case there are no additional $c$-dependent normalisation factors, leading to the level $N$ vacuum determinant formula up to a positive rational number given by
\begin{equation}
     \det\left(V_{\text{Vir},N}(c)\right) \sim \lim_{h \to 0} \frac{\det\, G^{(N)}_{\text{Vir}}(h;c)}{h^{p(N-1)}\det\, G^{(N-1)}_{\text{Vir}}(1+h;c)}~,
\end{equation}
where $G^{(N)}_{\text{Vir}}(h;c)$ and $G^{(N-1)}_{\text{Vir}}(1+h;c)$ are the Gram matrices for the level $N$ linear subspaces of the modules $M(h;c)$ and $M(1+h;c)$ at generic $h$ and $c$. After some simplification, this formula precisely matches the known level $N$ vacuum Kac determinant of the Virasoro algebra in \cite{Feigin:1982tg, gorelik2007simplicity} up to an overall ($c$-independent) positive number. 
\end{remark}

\section{\label{sec:preliminary_constraints}Graded unitarity structures for the \texorpdfstring{${\mathcal W}_3$}{W3} algebra}

Any vertex algebra associated to a unitary four-dimensional $\mN=2$ SCFT comes equipped with a graded unitary structure. This is a package of extra structures that consists of a (generally order-four) anti-linear conjugation $\rho$ as well as a (generally half-integral) nondegenerate good filtration $\mfR_\bullet$. The $\mfR$-filtrations on the VOAs associated with interacting four-dimensional SCFTs must obey additional conditions, and were dubbed \textit{interacting $\mfR$-filtrations (of four-dimensional type)} in \cite{ArabiArdehali:2025fad}. 

We are interested in cases where the underlying vertex algebra is taken to be the ${\mathcal W}_3$ algebra, in which case both of these structures simplify to some extent. Since ${\mathcal W}_3$ is a $\mathbb{Z}$-graded bosonic vertex algebra, the graded-unitary version of the spin-statistics relation \cite{Beem:2025guj} requires that the $\mfR$-filtration be integral. As a consequence, the conjugation $\rho$ must be an involution.

As $\rho$ is a vertex algebra automorphism, its action is completely determined by its action on the strong generators $T$ and $W$. Furthermore, on general grounds one always has $\rho(T)=T$ for the stress tensor. There are then only two conjugations that could conceivably play the role of the graded-unitary conjugation, namely $\rho(W)=W$ or $\rho(W)=-W$.\footnote{Note that in the more conventional setting of two-dimensional unitarity, one has an anti-linear involution $\phi$ and for the ${\mathcal W}_3$ algebra one takes $\phi(W)=-W$.} It will turn out that under the assumption of graded unitarity, the choice of which conjugation action on $W$ is realised is dictated by the $R$-charge assignment for $W$.

We recall that graded unitarity imposes constraints on the sign of the most singular term in the OPE of $\mO(z)$ and $\rho(\mO)(0)$ that correlate $\mO$'s position in the $\mfR$-filtration \cite{ArabiArdehali:2025fad}. In particular, writing the corresponding two point function as
\begin{equation} \label{basic}
    \langle \mathcal{O}(z) \rho(\mathcal{O})(0)\rangle = \frac{\kappa_\mathcal{O}}{z^{2h_\mathcal{O}}}~,
\end{equation}
graded unitarity requires
\begin{equation} \label{sign}
    \kappa_\mathcal{O} \propto (-1)^{h_\mathcal{O} - R_{\mathcal{O}}}~,
\end{equation}
with proportionality up to a positive constant. Here we assume that the $\mfR$-filtration has been refined to a grading by Gram--Schmidt, and $R_\mathcal{O}$ is the $R$-charge of $\mathcal{O}(z)$.

For a four-dimensional $\mfR$-filtration, one has that the stress tensor has $R$-charge $R_T = 1$, and so by \eqref{sign} applied to $T$ one has $c \leqslant 0$. This is the familiar VOA avatar of the statement that the central charge $c_{4d}$ of the parent SCFT must be non-negative. For the spin-three generator $W$ we have
\begin{equation}
    \langle W(z) W(0) \rangle \sim \frac{c/3}{z^6}~,
\end{equation}
whereas for graded unitarity we require\footnote{We may assume, without loss of generality, that $W$ has definite $R$-charge.}
\begin{equation}
    \langle W(z) \rho(W)(0)\rangle \sim \frac{\kappa_W}{z^{6}}~,\qquad\kappa_W \propto (-1)^{3 - R_W}~.
\end{equation}
Given $c \leqslant 0$, we are left with the following two possibilities,
\begin{itemize}
    \item $\rho(W) = W$ and $R_W \in 2 \mathbb{N}$,
    \item $\rho(W) = -W$ and $R_W \in 2\mathbb{N}+1$.
\end{itemize}

Finally, we note that without any assumptions beyond the statement that the $\mfR$-filtration is of four-dimensional interacting type, one has the further central charge constraints \cite{ArabiArdehali:2025fad}
\begin{equation} \label{prelimconst}
    c = 0\qquad\text{or}\qquad c \leqslant -\frac{22}{5}~.
\end{equation}
In the case $c=-22/5$, the ${\mathcal W}_3$ algebra degenerates to the $(2,5)$ Virasoro VOA, so in what follows we will take as given that $c<-22/5$.

As in the recent work of \cite{ArabiArdehali:2025fad}, the determination of the full $\mfR$-filtration from first principles will be beyond our abilities in this work. Instead, we will \emph{assume} that the $\mfR$-filtration is a \emph{weight-based filtration} with generating subspace $\{T,W\}$. By this, we mean that we will define 
\begin{equation}
    \mfR_p{\mathcal W}_3 = {\rm span}\{L_{-2-n_1}\ldots L_{-2-n_i} W_{-3-m_1}\ldots W_{-3-m_j} \Omega~,~~i+jR_W \leqslant p\}~,
\end{equation}
where the stress tensor is assigned $R_T = 1$ as above and the $R$-charge of $W$ is fixed to some positive integer value.\footnote{For the filtration to be of four-dimensional type, we must have $1\leqslant R_W\leqslant 3$, and for it to be of interacting type we must have $2\leqslant R_W\leqslant 3$, but it will not be necessary to impose these in what follows. In the actual Argyres--Douglas models giving rise to ${\mathcal W}_3$ VOAs one will have $R_W=2$.} 

Note that the choice for $R_W$ to be even or odd must be accompanied by the choice of conjugation $\rho(W)=W$ or $\rho(W)=-W$, respectively. Then the sign rule \eqref{basic} and \eqref{sign} for two-point functions of operators involving $W$ currents are actually \emph{independent} of this choice (as extra signs in \eqref{basic} coming from the action of $\rho$ are accounted for by extra signs coming from odd $R$-charges in \eqref{sign}). With this generality in mind, for the sake of concreteness in our calculations we will henceforth specialise to the choice $R_W=2$. This is the choice of $\mfR$-filtration which is expected to be pertinent to the $(A_2,A_q)$ Argyres--Douglas theories.

\section{\label{sec:classification}Constraints from graded unitarity}

We will now examine the constraints imposed on the central charge of the ${\mathcal W}_3$ algebra under the assumption of graded unitarity and an $\mfR$-filtration of the type discussed in the previous section. Our strategy will be the same as that applied in \cite{ArabiArdehali:2025fad}: we require that the Kac determinant \eqref{rationalc} has the same sign (at every level) as that which is predicted by graded unitarity. 

Recall that the graded unitarity constraint \eqref{sign} is a constraint on the signs of two-point functions between operators and their $\rho$-conjugates,
\begin{equation}
    \text{sign}\left(z^{2h_\mO}\langle\rho(\mO)(z)\mO(0)\rangle\right) = (-1)^{h_\mO-R_\mO}~.
\end{equation}
This translates into a sign constraint for the Kac determinant, which requires considering the relation between signs of Shapovalov norms and two-point functions,
\begin{equation}
   \text{sign}\left( \langle \mO | \mO \rangle \right) = (-1)^{h_\mO} \text{sign} \left( \lim_{z \to \infty} z^{2 h_\mO} \langle \phi(\mO)(z) \mO(0) \rangle \right)~,
\end{equation}
where $|\mO\rangle$ is the state corresponding to the operator $\mO(z)$ via the state operator correspondence, and $\phi$ is an anti-linear involution whose definition is implicit in the rules for hermitian conjugation for modes of $\mO$. In the case of the ${\mathcal W}_3$ algebra, the condition $W_n^\dagger = W_{-n}$ and $L_n^\dagger = L_{-n}$ translates to the involution action $\phi(T) = T$ and $\phi(W) = -W$. (For more details on these sign relations, see, \emph{e.g.}, Appendix B of \cite{ArabiArdehali:2025fad}.) 

The sign constraints of graded unitarity are then easily understood by introducing the (linear) $\mathbb{Z}_2$ automorphism on the ${\mathcal W}_3$ algebra that sends $(T,W)\to (T,-W)$. One can choose a basis for ${\mathcal W}_3$ of states with definite $R$-charge (by refining the $\mfR$-filtration by Gram-Schmidt) and definite parity $\sigma(\mO)$ under this $\mathbb{Z}_2$ automorphism.\footnote{This parity counts the parity of the number of $W$ operators appearing in an expression for the operator as a normally ordered product of $T$'s and $W$'s. Since the $\mathbb{Z}_2$ action is an automorphism, the $\mfR$-filtration can be diagonalised independently for even and odd parity subsectors.} Chasing the various sign relations, graded unitarity then requires the sign of the Shapovalov norm of a state with fixed $R$-charge and $\sigma$-parity to be as follows,
\begin{equation}\label{gramrelation}
    \text{sign}\left( \langle \mO | \mO \rangle \right) = \begin{cases}
        (-1)^{R_\mO - \sigma(\mO)} & \qquad \text{if } \quad \rho(W) = W \quad \text{and} \quad R_W \in 2 \mathbb{N}~,\\
        (-1)^{R_\mO} & \qquad \text{if } \quad \rho(W) = -W \quad \text{and} \quad R_W \in 2 \mathbb{N}+1~.  
    \end{cases}
\end{equation}
In fact, these give rise to the same sign constraint on the Shapovalov norm $\langle \mO | \mO \rangle$. Hence, without loss of generality, we consider the case with $R_W = 2$, in which case we have the following requirement for the level-$N$ vacuum Kac determinant, 
\begin{equation}\label{mainconst}
    \text{sign}(\det\, V_N(c)) =
    \prod_{\substack{
        \{\lambda_1, \lambda_2\} \vdash N \\
        1 \notin \lambda_1,\\\{1,2\} \notin \lambda_2}}(-1)^{R_\lambda - \sigma_\lambda}~,
\end{equation}
where $\lambda = \{\lambda_1, \lambda_2 \}$ is a two-coloured partition of $N$. Each such  partition $\lambda$ corresponds to a specific PBW-ordered state $L_{-a_1} \ldots L_{-a_i} W_{-b_1} \ldots W_{-b_j} |0\rangle$ with $\lambda_1 = \{a_1, \ldots, a_i\}$ and $\lambda_2 = \{b_1, \ldots, b_j\}$. For such a state, $R_{\lambda}=i+2j$ and $\sigma_\lambda=j~{\rm mod}~2$.

Equation \eqref{mainconst} will be our window into the constraints imposed by graded unitarity. Even though it is a small subset of the state-level sign constraints that are required in order for graded unitarity to hold, we will see that this determinant-level condition is sufficient to eliminate all central charges other than the ``boundary'' $(3,q)$ levels,
\begin{equation}
    c_{3,q} = 2 - 8 \frac{(q-3)^2}{q}~,
\end{equation}
with $q > 3$ and $(3,q) = 1$. 

\subsection{\label{lowlevel}Explicit computations of low-level vacuum Kac determinants}

As a warm up, we consider graded unitarity constraints at low levels by a hands-on analysis. As reviewed above, up to level four, the constraints of graded unitarity are independent of the specified $\mfR$-filtration, and are identical to those identified in the purely Virasoro case. In particular, assuming an interacting $\mfR$-filtration, one has
\begin{equation}
    c = 0 \qquad \text{or} \qquad c \leqslant -\frac{22}{5}~.
\end{equation}
Following this, we use the $\mfR$-filtration choice of the previous section to impose graded unitarity constraints on central charges by examining low-level (levels $N = 5$ to $N = 10$) vacuum Kac determinants.

At level 5, the vacuum Kac matrix is the Gram matrix for the Shapovalov form evaluated on the following basis states,
\begin{equation}
    \mathcal{B}_5 = \left\{ L_{-2}W_{-3}|0,0\rangle~,~W_{-5}|0,0\rangle~,~ L_{-3}L_{-2}|0,0\rangle~,~L_{-5}|0,0\rangle\right\}~.
\end{equation}
These signs $(-1)^{R_\lambda - \sigma_\lambda}$ for these basis states are given by $\{1, -1, 1, -1\}$, so graded unitarity implies that the level-five vacuum Kac determinant should be positive. Computing by hand, this determinant is given by
\begin{equation}
    \det\, V_5(c)  \sim \textcolor{blue}{( 114 + 7c)}  (22 + 5c)  c^4~.
\end{equation}
where proportionality is, as usual, up to a positive real constant. Graded unitarity then requires
\begin{equation}
    ( 114 + 7c)(22 + 5c)c^4 \geqslant 0~.
\end{equation}
Combining this with the preliminary central charge constraints, we obtain
\begin{equation}
    \textcolor{blue}{c \leqslant c_{3,7}=-\frac{114}{7}}  \qquad \text{or} \qquad c = \left\{-\frac{22}{5}~, 0 \right\}~.
\end{equation}
In other words, the interval between $c_{3,5}=-\frac{22}{5}$ and $c_{3,7}=-\frac{114}{7}$ is excluded.

By a straightforward analysis of a PBW-like basis of states, at level six graded unitarity implies that the vacuum Kac determinant should be negative. The level-six vacuum Kac determinant for the ${\mathcal W}_3$ algebra is then given by
\begin{equation}
    \det\, V_6(c) \sim \textcolor{blue}{(23 + c)} (114 + 7 c)^2 (22 + 5 c) (2 + c)^2 c^8  (-4 + 5 c)\leqslant 0~.
\end{equation}
Combining this constraint with the lower-level constraints, one has the further restriction,
\begin{equation}
    \textcolor{blue}{c \leqslant c_{3,8}=-23} \qquad \text{or} \qquad c = \left\{-\frac{114}{7}~, -\frac{22}{5}~, 0 \right\}~.
\end{equation}

To see the pattern of constraints, it is instructive to additionally consider the level-seven case as well. Here, the Kac determinant takes the form 
\begin{equation}
    \det\, V_7(c) \sim \textcolor{blue}{(23 + c)^2} (114 + 7 c)^3 (22 + 5 c) (2 + c)^2 c^{10} (-4 + 5 c)^2~,  
\end{equation}
and graded unitarity requires this to be positive. This then gives no new constraints compared to the lower-level restrictions.

We give several more explicit low-level computations of the vacuum Kac determinant of the ${\mathcal W}_3$ algebra below;
\begin{equation}
\begin{split}
    &\det\, V_8(c)  \sim \textcolor{blue}{(186 + 5 c)}(23 + c)^3 (114 + 7 c)^6 (22 + 5 c) (2 + c)^6 c^{17}(-4 + 5 c)^5~,\\
    &\det\, V_9(c)  \sim \frac{\textcolor{blue}{(490 + 11 c)} (186 + 5 c)^2 (23 + c)^6 (114 +  7 c)^{10}(2 + c)^{10} c^{24}(4 - 5 c)^8}{22 + 5c}~,\\
    &\det\, V_{10}(c) \sim \frac{\textcolor{blue}{(490 + 11 c)^2} (186 + 5 c)^3 (23 + c)^{10} (114 + 7 c)^{16} (40 + 7 c) (2 + c)^{18} c^{36} (4 - 5 c)^{14}}{(22+5c)^2}~,
\end{split}
\end{equation}
and we summarise the constraints obtained from the sign requirements on these determinants in Table \ref{tab:ccconst}.
\begin{table}[t]
\centering
\renewcommand{\arraystretch}{1.6}
\begin{tabular}{|c|l|}
\hline
\textbf{Level} & \textbf{Central charge constraints} \\
\hline
2 & \( c \leqslant 0 \) \\
\hline
3 & \( c \leqslant 0 \) \\
\hline
4 & \( c \in \{ 0 \} \cup \left(-\infty, \textcolor{blue}{-\frac{22}{5}}\right] \) \\
\hline
5 & \( c \in \left\{-\frac{22}{5},  0 \right\} \cup \left(-\infty, \textcolor{blue}{-\frac{114}{7}}\right] \) \\
\hline
6 & \( c \in \left\{-\frac{114}{7}, -\frac{22}{5}, 0 \right\} \cup \left(-\infty, \textcolor{blue}{-23}\right] \) \\
\hline
7 & \( c \in \left\{-\frac{114}{7}, -\frac{22}{5}, 0 \right\} \cup \left(-\infty, \textcolor{blue}{-23}\right] \) \\
\hline
8 & \( c \in \left\{-23, -\frac{114}{7}, -\frac{22}{5}, 0 \right\} \cup \left(-\infty, \textcolor{blue}{-\frac{186}{5}}\right] \) \\
\hline
9 & \( c \in \left\{-\frac{186}{5}, -23 , -\frac{114}{7}, 0 \right\} \cup \left(-\infty, \textcolor{blue}{-\frac{490}{11}}\right] \) \\
\hline
10 & \( c \in \left\{-\frac{186}{5}, -23 , -\frac{114}{7},  0  \right\} \cup \left(-\infty, \textcolor{blue}{-\frac{490}{11}}\right] \) \\
\hline
\end{tabular}
\caption{\label{tab:ccconst}Central charge constraints from graded unitarity at levels $N=2,\ldots,10$}.
\end{table}

We see that at levels seven and ten, no new constraints arise. This is a general feature of the Kac determinant constraints, and we will prove in the next subsections that for $N\geqslant 5$, no new constraints arise at level $N = 1~{\rm mod}~3$.\footnote{It is a technical, but perhaps interesting, point that this pattern does continue to hold for $N < 5$. In particular, there is a new constraint at level four, but not at level three. This  is because, as we will see later, these bounds depend on the structure of zeros of \eqref{rationalc}, and at level three, a factor of $(c - c_{3,5})$ coming from the product is canceled by an explicit $1/(22+5c)$ in the prefactor. This effectively ``hides'' the constraint that should arise at level three, and delays its appearance until level four.} 

After observing the pattern of the constraints up to level ten, one might expect that at higher levels, it would continue to be the case that $N=1~({\rm mod}~3)$ there are no new constraints, while for $N = 2~({\rm mod}~3)$ a new constraint eliminates the interval $(c_{3,N+2}, c_{3,N})$ and for $N = 0~({\rm mod}~3)$ the new constraint eliminates the interval $(c_{3,N+2}, c_{3,N+1})$.

The subsections below are dedicated to establishing that this is exactly what happens at general level. Eventually, this implies that graded unitarity rules out all values for the central charge other than: 
\begin{equation}
    c = \{c_{3,q} = 2 - 8 \frac{(q-3)^2}{q} : \qquad q > 3 \qquad \& \qquad (3,q) = 1\}~.
\end{equation}
Of course, these are the central charges of the $(3,q)$ ${\mathcal W}_3$ minimal model VOAs, which are precisely those associated with the $(A_2, A_{q-4})$ Argyres-Douglas theories \cite{Cordova:2015nma}. 

Our strategy will make extensive use of the closed-form expression for the vacuum determinant derived in \eqref{rationalc}. We will show that the most negative zero of the level $N$ vacuum Kac determinant is at $c=c_{3,N+2}$ for $N \neq 1\ {\rm mod}\ 3$ and has order one, while for $N = 1\ {\rm mod}\ 3$ it is at $c=c_{3,N+1}$ and has order two. Moreover, when $N\neq 1\ {\rm mod}\ 3$, the second-to-most negative zero is at $c=c_{3,N+1}$ or $c=c_{3,N}$ depending on whether $N = 0\ {\rm mod}\ 3$ or $N = 2\ \text{mod}\ 3$, respectively. This will then establish the desired result so long as the sign of the level $N$ vacuum Kac determinant is consistent with graded unitarity for \emph{large negative central charge}. It is this last point to which we turn next. 
 
\subsection{\label{subsec:large_neg_c}Large negative central charge}

Consider the level $N$ vacuum Kac determinant det $(V_N(c))$. It is a rational function of $c$ of some finite degree $d(N)$ which can be computed as follows,
\begin{equation}\label{degree}
    d(N) = \sum_{\substack{
        \{\lambda_1, \lambda_2\} \vdash N\\
        1 \notin \lambda_1\\
        \{1,2\} \notin \lambda_2
    }} l(\lambda)~,
\end{equation}
where $l(\lambda)$ is the length of the two-coloured partition $\lambda$. This follows from the fact that the maximum power of $c$ obtained while computing the matrix element of the state corresponding to $\lambda$, 
$$
\langle0|W_{b_j} \ldots W_{b_1} L_{a_i} \ldots L_{a_1} L_{-a_1} \ldots L_{-a_i} W_{-b_1} \ldots W_{-b_j}|0 \rangle~,
$$
arises precisely when one isolates the contribution of the order-$c$ term in the commutation relations $[L_n, L_{-n}]$ and $[W_n, W_{-n}]$, giving $c^{i+j}$.\footnote{It might appear that the non-linear terms in the $WW$ commutators could lead to a higher power of $c$, but each non-linear term itself comes with a factor of $\frac{1}{22+5c}$, so the maximum power cannot actually be increased.} Hence, for sufficiently large negative $c$, the sign of the vacuum Kac determinant given by
\begin{equation}
    \text{sign}(\det\, V_N(c)) = (-1)^{d(N)}~.
\end{equation}
That this precisely matches the right hand side of equation \eqref{mainconst} is now more or less immediate. Indeed, for a specific two-coloured partition $\lambda$, one has
\begin{equation}
    R_\lambda = i + 2j~, \qquad \sigma_\lambda = j~,
\end{equation}
So the right-hand side of \eqref{mainconst}, which is the prediction of graded unitarity for the sign of the vacuum determinant, is given by: 
\begin{equation}
    \text{sign}_{\text{gr.unit.}} \equiv \prod_{\substack{
        \{\lambda_1, \lambda_2\} \vdash N \\
        1 \notin \lambda_1,\\\{1,2\} \notin \lambda_2}} (-1)^{R_\lambda - \sigma_\lambda} = \prod_{\substack{
        \{\lambda_1, \lambda_2\} \vdash N \\
        1 \notin \lambda_1,\, \{1,2\} \notin \lambda_2
    }} (-1)^{i+j} = \prod_{\substack{
        \{\lambda_1, \lambda_2\} \vdash N \\
        1 \notin \lambda_1,\, \{1,2\} \notin \lambda_2
    }} (-1)^{l(\lambda)}~.
\end{equation}
This is exactly what we have for the Kac determinant in the large negative central charge regime.

We can therefore turn to the next issue of analysing the structure of zeros of the Kac determinant in order to establish the predicted constraints from the previous subsection. 

\subsection{\label{subsec:kac_zeros}Analysis of zeros of the vacuum Kac determinant}

The first obvious point that can be seen from the closed-form expression of the vacuum Kac determinant \eqref{rationalc} is that the only possible zeros of the level-$N$ vacuum Kac determinant are at values
\begin{equation} \label{assump}
    c = c_{p,q} = 2 - 24\frac{(p-q)^2}{pq}~,\qquad q>p\geqslant 2~,\quad (p,q)=1~.
\end{equation}
Hence, the zeros essentially fall into the following two classes:
\begin{itemize}
    \item $c = c_{p,q}$,~~$(p,q) = 1$ \text{and} $q>p\geqslant 3$~.
    \item $c = c_{2,q}$,~~$(2,q) = 1$ \text{and} $q\geqslant3$~.
\end{itemize}
It is useful to divide the set of zeros into these two classes because the manner of appearance of the two sets of zeros is qualitatively different (as will become evident from the exponents $E_{p,q}^{(N)}$). We will first show that the most negative and the second most negative zeros always fall in the first class with $p = 3$. In summary, we will find the following structure:
\begin{itemize}
    \item At level $N = 0 \ \text{mod} \ 3$, the most negative zero at level $N$ is given by $c = c_{3,N+2}$, and this zero always occurs with multiplicity one. Moreover, the second-to-most negative zero is given by $c = c_{3,N+1}$.
    \item At level $N = 1 \ \text{mod} \ 3$, the most negative zero at level $N$ is given by $c = c_{3,N+1}$, and this zero always occurs with multiplicity two.
    \item At level $N = 2 \ \text{mod} \ 3$, the most negative zero at level $N$ is given by $c = c_{3,N+2}$ (with unit multiplicity), and the second-to-most negative zero is given by $c = c_{3,N}$.
\end{itemize}
These facts immediately lead to the previously advertised constraints for the central charge. 

It is important to identify the lowest level $N$ at which a particular $c=c_{p,q}$ can occur as a zero of the Kac determinant. A straightforward analysis of the exponents \eqref{exponent} reveals that for certain levels $N$, the central charge $c_{p,q}$ with $(p,q) = 1$ cannot be a zero of the level $N$ Kac determinant (\emph{i.e.}, where the exponents $E^{(N)}_{p,q}$ vanish). In particular, one sees that
\begin{itemize} \label{zeroexp}
    \item $E^{(N)}_{2,q} = 0$ for all $N < q-1$.
    \item $E^{(N)}_{p,q} = 0$ for all $N < (p-2)(q-2)$.
\end{itemize}
These elementary observations show that at level $N$, there can only be zeros at $c = c_{2,q}$ for $q \leqslant N+1$. Having seen the universal graded unitarity constraints that are visible up to level four, we will restrict ourselves to $N \geqslant 5$. 

Recalling the expression for the $c_{p,q}$ central charges,
\begin{equation}
    c_{p,q} = 2 - 24 \frac{(p-q)^2}{p q} = 2 - 24 \left(\frac{p}{q} + \frac{q}{p} - 2 \right)~,
\end{equation}
we can make some general observations about the possible $c_{3,q}$ zeros of the level-$N$ Kac determinant.
\begin{itemize}
    \item The most negative possible zero of the form $c_{3,q}$ is $c_{3,N+2}$ for $N \neq 1\ \text{mod}\ 3$ and $c_{3,N+1}$ for $N = 1\ \text{mod}\ 3$.
    \item The second most negative possible zero of the form $c_{3,q}$ is given by $c_{3,N+1}$ for $N = 0 \ \text{mod} \ 3$, and $c_{3,N}$ for $N = 2 \ \text{mod} \ 3$.
\end{itemize}
Now $c_{2,q} \leqslant c_{3,N+1}$ (or $c_{3,N}$) if and only if $q \geqslant \left\lceil \frac{2 (N+1)}{3} \right\rceil$ (or $\left\lceil \frac{2 N}{3} \right\rceil$). Combining this with the previous condition that $q \leqslant N+1$, we find that the possible level $N$ zeros of the form $c = c_{2,q}$ that are more negative than $c_{3,N}$ must satisfy
\begin{align} \label{qconst}
    \begin{aligned}
        &\left\lceil \frac{2 (N+1)}{3} \right\rceil &&\leqslant q \leqslant N+1 && \text{for}\ \ N \neq 2 \ \text{mod} \ 3~,\\ 
        &\left\lceil \frac{2 N}{3} \right\rceil &&\leqslant q \leqslant N+1 && \text{for}\ \ N = 2 \ \text{mod} \ 3~.
    \end{aligned}
\end{align}
Hence, our first step will be to show that the level $N$ vacuum Kac determinant does not have zeros of the form $c=c_{2,q}$ with $q$ satisfying \eqref{qconst}. Indeed, when $q$ lies in this range, the expression for the exponent of $(c-c_{2,q})$ in the vacuum Kac determinant \eqref{rationalc} simplifies and can be seen to vanish,
\begin{equation}
    \begin{split}
    E_{2,q}^{(N)} &= 2P_2(N - (q-1)) - 2 P_2(N - 1 - (q-2)) - 2P_2(N-1-(q+1))\\
    &+ 2P_2(N-3-(q-1)) + 2P_2(N-3-(q+2)) - 2P_2(N-4-(q+1))= 0~.
    \end{split}
\end{equation}
which implies that the level $N$ vacuum Kac determinant does not have zeros at $c=c_{2,q}$ with $q \geqslant \lceil \frac{2N}{3} \rceil$. Now we turn to the case of zeros of the form $c = c_{p,q}$ with $p \geqslant 4$. 

\paragraph{Zeros at $c = c_{p,q}$ with $p \geqslant 4$ and $(p,q) = 1$.} 
Without loss of generality, consider $q > p$. Since $x + \frac{1}{x}$ is an increasing function for $x > 1$, for a fixed $p$, $c_{p,q}$ is the least when $q$ is maximum. 

On the other hand, from \eqref{zeroexp}, one sees that for a given level $N \geqslant 4$, the values of $q$ for which the vacuum Kac determinant can possibly have a zero $c_{p,q}$ for $p \geqslant 4$ satisfy
\begin{align}
    q \leqslant \frac{N}{p-2} + 2 \leqslant N~.
\end{align}
This implies that for $p \geqslant 4$,
\begin{equation} 
    \frac{q}{p} \leqslant \frac{N}{4} < \frac{N}{3} \Longrightarrow 
    \begin{cases}
        c_{p,q} > c_{3,N+1} \qquad \text{if} \ \ N \neq 2 \ \text{mod} \ 3~.\\
        c_{p,q} > c_{3,N} \quad\qquad \text{if} \ \ N = 2 \ \text{mod} \ 3~.
    \end{cases}
\end{equation}
This shows that at level $N$, a possible zero at $c=c_{p,q}$, with $p\geqslant 4$ and $(p,q) = 1$ is always greater (less negative) than $c_{3,N+1}$ (or $c_{3,N}$); these were the second most negative possible zeros of the form $c_{3,q}$.

\paragraph{The smallest zeros $c = c_{3,q}$.}
The previous analysis of zeros completes most of the argument, and establishes that the most negative (possible) zeroes of the level $N$ vacuum Kac determinant are indeed at $c = c_{3,N+2}$ or $c = c_{3,N+1}$, and the second most negative (possible) zeroes are at $c = c_{3,N+1}$ or $c = c_{3,N}$, depending on the coprimality condition $(N,3) = 1$. The multiplicities of these are obtained in a straightforward manner from the exponents of $(c-c_{3,q})$, which can be computed directly and give
\begin{equation}
    \begin{split}
    &E_{3,N+2}^{(N)} = P_2(0) = 1 \qquad \text{for} \ N \neq 1 \ \text{mod} \ 3~, \\
    &E_{3,N+1}^{(N)} = P_2(1) = 2 \qquad \text{for} \ N = 1 \ \text{mod} \ 3~.
    \end{split}
\end{equation}
This proves that the structure of zeros of the level $N$ vacuum Kac determinant of the ${\mathcal W}_3$ algebra is as described above,
\begin{itemize}
    \item For $N \neq 1 \ \text{mod} \ 3$, $c = c_{3,N+2}$ is the most negative zero and has multiplicity one, and $c = c_{3,N+1}$ (for $N = 0 \ \text{mod} \ 3$) or $c = c_{3,N}$ (for $N = 2 \ \text{mod} \ 3$) is the second to most negative zero.
    \item For $N = 1 \ \text{mod} \ 3$, $c = c_{3,N+1}$ is the most negative zero and has multiplicity two. 
\end{itemize} 
These results now allow us to extract constraints on the central charges compatible with graded unitarity

\subsection{Central charge constraints}

The conclusion of the analysis in the previous subsection is that the structure of the level $N$ vacuum Kac determinant takes the following form: 

\underline{\textbf{Case 1: }$N \neq 1 \ \text{mod} \ 3$}:
\begin{equation} \label{KacDformcp}
\begin{split}
    &\det V_N(c) \sim \textcolor{blue}{(c - c_{3,N+2})^1} (c-c_{3,N+1})^{\#_1} (\ldots\ldots\ldots) \qquad N = 0 \ \text{mod} \ 3~,\\
    &\det V_N(c) \sim \textcolor{blue}{(c - c_{3,N+2})^1} (c-c_{3,N})^{\#_2} (\ldots\ldots\ldots) \qquad N = 2 \ \text{mod} \ 3~,
    \end{split}
\end{equation}
where $(\ldots\ldots\ldots)$ denotes zeros of the vacuum Kac determinant that are strictly greater (less negative) than $c_{3,N+1}$ (or $c_{3,N}$). $\#_1$ and $\#_2$ are positive integers.

\underline{\textbf{Case 2: }$N = 1\text{ mod }3$}:
\begin{align} \label{KacDformnotcp}
    \det V_N(c) \sim \textcolor{blue}{(c - c_{3,N+1})^2}(\ldots\ldots\ldots)~,
\end{align}
where again $(\ldots\ldots\ldots)$ correspond to other zeros of the vacuum Kac determinant that are strictly greater than $c_{3,N+1}$. 

The all-level constraints of graded unitarity (at the level of the Kac determinant) now follow almost immediately, given the previously established fact that the sign of the vacuum determinant for large negative central charge is consistent with the sign predicted by graded unitarity. 

\begin{figure}[t]
\centering

\begin{subfigure}[b]{0.95\textwidth}
\centering
\hspace*{-0.5cm}
\begin{tikzpicture}[scale=1.2]
\begin{scope}[xshift=-5cm]

    \draw[very thick, black] (-1,0) -- (0,0);          
    \draw[very thick, blue] (0,0) -- (2,0);            
    \draw[very thick, blue, dashed] (2,0) -- (4,0);    
    \draw[very thick, blue, dashed] (4,0) -- (5,0);     
    \draw[very thick, blue, dashed] (7,0) -- (8,0); 
    \draw[very thick, blue, dashed] (8,0) -- (10,0);  
    \draw[very thick, blue, dashed] (10,0) -- (11,0);  

    \foreach \i/\x in {2/0, 1/2} {
        \fill[red] (\x,0) circle (3pt);
        \node[below] at (\x,-0.1) {$c_{3,N+\i}$};
    }
    \fill[red] (4,0) circle (3pt); 
    \node[below] at (4, -0.1) {$c_{3,N-1}$};
    \fill[red] (8,0) circle (3pt); 
    \node[below] at (8, -0.1) {$c_{3,7} = -\frac{114}{7}$};
    \fill[red] (10,0) circle (3pt); 
    \node[below] at (10, -0.1) {$c_{3,4} = 0$};
 
    \node at (-1.4,0) {\Large$\ldots\ldots$}; 
    \node at (6,0) {\Large\textcolor{blue}{$\ldots\ldots\ldots\ldots$}};


\end{scope}
\end{tikzpicture}
\caption{Level $N = 0 \mod 3$: Thick blue segment corresponds to new excluded region. Red dots indicate level $N$ Kac determinant zeros of the form $c_{3,q}$ with $(3,q) = 1$.}
\end{subfigure}

\vspace{1cm}

\begin{subfigure}[b]{0.95\textwidth}
\centering
\hspace*{-0.5cm}
\begin{tikzpicture}[scale=1.2]
\begin{scope}[xshift=-5cm]

    \draw[very thick, black] (-1,0) -- (0,0);          
    \draw[very thick, blue, dashed] (0,0) -- (2,0);    
    \draw[very thick, blue, dashed] (2,0) -- (4,0);    
    \draw[very thick, blue, dashed] (4,0) -- (5,0);     
    \draw[very thick, blue, dashed] (7,0) -- (8,0); 
    \draw[very thick, blue, dashed] (8,0) -- (10,0);  
    \draw[very thick, blue, dashed] (10,0) -- (11,0);  

    \fill[red] (0,0) circle (3pt); 
    \node[below] at (0, -0.1) {$c_{3,N+1}$};
    \fill[red] (2,0) circle (3pt); 
    \node[below] at (2, -0.1) {$c_{3,N}$};
    \fill[red] (4,0) circle (3pt); 
    \node[below] at (4, -0.1) {$c_{3,N-1}$};
    \fill[red] (8,0) circle (3pt); 
    \node[below] at (8, -0.1) {$c_{3,7} = -\frac{114}{7}$};
    \fill[red] (10,0) circle (3pt); 
    \node[below] at (10, -0.1) {$c_{3,4} = 0$}; 

    \node at (-1.4,0) {\Large$\ldots\ldots$}; 
    \node at (6,0) {\Large\textcolor{blue}{$\ldots\ldots\ldots\ldots$}};


\end{scope}
\end{tikzpicture}
\caption{Level $N = 1 \mod 3$. No new excluded region at level $N$ as the entire dashed blue line is already excluded by graded unitarity constraints at level $N-1$.}
\end{subfigure}

\caption{Central charge constraints at levels $N = 0 \text{ mod }3$ and $N = 1 \text{ mod } 3$. Dashed blue lines indicate regions excluded at earlier levels. Black lines and red dots indicate allowed central charges not ruled out by graded unitarity constraints at level $N$.}
\label{fig:combined_c_constraints}
\end{figure}
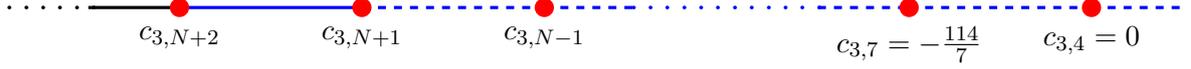
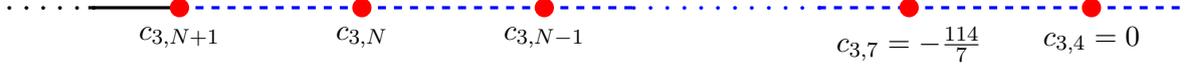

In particular, let us first consider $N \neq 1 \text{ mod }3$. From \eqref{KacDformcp} we see that the vacuum Kac determinant flips sign from that predicted by graded unitarity (value for large negative $c$) to the wrong sign at $c = c_{3,N+2}$, with the next zero being at $c_{3,N+1}$ or $c_{3,N}$. Hence, at level $N$, graded unitarity rules out the possibility of $c \in (c_{3,N+2}, c_{3,N+1})$ or $c \in (c_{3,N+2}, c_{3,N})$ (depending on whether $N = 0 \text{ mod }3$ or $N = 2 \text{ mod }3$). 

Turning to the $N = 1 \text{ mod }3$ case, note that the most negative zero is $c = c_{3,N+1}$, and all the $c_{3,q}$ zeros (with $N+1 \geqslant q > 3$ and $(3,q) = 1$) of the level $N-1$ vacuum Kac determinant are also zeros of the level $N$ vacuum determinant. With the level $N-1 = 0 \text{ mod }3$ graded unitarity constraints already ruling out the region $(c_{3,N-1+2}, c_{3,N-1+1})$, we do not get any new constraints from graded unitarity at level $N = 1 \text{ mod }3$. For clarity, we have succinctly shown the case for $N = 0\text{ mod }3$ and $N = 1\text{ mod }3$ in Figure \ref{fig:combined_c_constraints}. 

To conclude, working towards a classification of graded unitary ${\mathcal W}_3$ algebras, we have shown that for a very large class of $\mfR$-filtrations (namely, all good, increasing, weight-based filtrations generated by $T$ and $W$), the (all-level) Kac-determinant-based graded unitarity constraints rule out all possible values for central charges other than the $(3,q)$ ${\mathcal W}_3$ minimal model central charges. These ${\mathcal W}_3$ algebra central charges are precisely the ones obtained by a Drinfel'd--Sokolov reduction of the boundary-admissible-level $\mathfrak{sl}_3$ affine current algebras at level $k = -3+\frac{3}{q}$, which were singled out by an analogous graded unitarity analysis in \cite{ArabiArdehali:2025fad}. These are the vertex algebras that appear as the associated VOAs of the $(A_2, A_{q-4})$ Argyres-Douglas SCFTs via the SCFT/VOA correspondence.


\acknowledgments

The authors thank Tomoyuki Arakawa, Dylan Butson, Gurbir Dhillon, Niklas Garner, Leonardo Rastelli, and Ben Webster for helpful discussions on this and related work. The work of CB was supported in part by grant \#494786 from the Simons Foundation, by ERC Consolidator Grant \#864828 ``Algebraic Foundations of Supersymmetric Quantum Field Theory'' (SCFTAlg), and by the STFC consolidated grant ST/X000761/1. The work of HK was supported by the Clarendon Fund Scholarship and the Eddie Dinshaw Scholarship at Balliol College.


\vfill\eject

\appendix


\section{\label{app}Kac determinant computation details} 

The Verma resolution \eqref{resol} and the related singular vector embedding diagram \eqref{fig:submod-structure} describe the structure of the vacuum Verma module. By a similar analysis (or alternatively from the corresponding structure of Vermas and simple modules in the regular block of BGG category $\mathcal{O}$ for $\mathfrak{g}=\mathfrak{sl}_3$) one obtains the structure of singular vectors in the Verma modules with highest weights $(1,w_\pm)$ and $(3,\lambda_\pm)$, and equivalently the resolution by Verma modules of the corresponding simple quotient modules (\emph{i.e.}, the other composition factors of the vacuum Verma module). Indeed, we have the following such resolutions.
The finite resolutions for the irreducible quotients $L(1,w_\pm,c)$ and $L(3,\lambda_\pm,c)$ are given by: 
\begin{equation}
    \begin{tikzpicture}[baseline=(current  bounding  box.center),
        node distance=8mm and 8mm,
        every node/.style={inner sep=2pt}
    ]

    \node (Zero1) {$0$};
    \node[right=of Zero1] (M3) {$\scriptstyle{M(4,0;c)}$};
    \node[right=of M3] (M2) {$\substack{M(3,\lambda_+;c)\\ \oplus\\M(3,\lambda_-;c)}$};
    \node[right=of M2] (M1) {$\scriptstyle{M(1,w_\pm;c)}$};;
    \node[right=of M1] (L1) {$\scriptstyle{L(1,w_\pm,c)}$};
    \node[right=of L1] (Zero) {$0$};

    \draw[->] (Zero1) -- (M3) node[midway, above] {$\scriptstyle{d_4}$};
    \draw[->] (M3) -- (M2) node[midway, above] {$\scriptstyle{d_3}$};
    \draw[->] (M2) -- (M1) node[midway, above] {$\scriptstyle{d_2}$};
    \draw[->] (M1) -- (L1) node[midway, above] {$\scriptstyle{d_1}$};
    \draw[->] (L1) -- (Zero) node[midway, above] {$\scriptstyle{d_0}$};
    
    \end{tikzpicture}
\end{equation}
\begin{equation}
    \begin{tikzpicture}[baseline=(current  bounding  box.center),
        node distance=8mm and 8mm,
        every node/.style={inner sep=2pt}
    ]

    \node (Zero1) {$0$};
    \node[right=of Zero1] (M2) {$\scriptstyle{M(4,0;c)}$};
    \node[right=of M2] (M1) {$\scriptstyle{M(3,\lambda_\pm,c)}$};;
    \node[right=of M1] (L3) {$\scriptstyle{L(3,\lambda_\pm,c)}$};
    \node[right=of L3] (Zero) {$0$};

    \draw[->] (Zero1) -- (M2) node[midway, above] {$\scriptstyle{d_3}$};
    \draw[->] (M2) -- (M1) node[midway, above] {$\scriptstyle{d_2}$};
    \draw[->] (M1) -- (L3) node[midway, above] {$\scriptstyle{d_1}$};
    \draw[->] (L3) -- (Zero) node[midway, above] {$\scriptstyle{d_0}$};
    
    \end{tikzpicture}
\end{equation}
These will in turn be used, starting with \eqref{jantzendet}, to ultimately express the vacuum Kac determinant \eqref{vacdet} in terms of Verma module determinants.

Let us introduce some notation. Throughout this appendix, the choice of central charge $c$ will be left implicit in the names of ${\mathcal W}_3$ modules. The irreducible quotients of Verma modules $M(1,w_\pm)$ and $M(3,\lambda_\pm)$ at generic $c$ will be denoted by $L(1,w_\pm)$ and $L(3,\lambda_\pm)$. The Gram matrices for the conformal weight $N$ linear subspace of these irreducible quotients will be denoted by $L_{N-1}(1, w_\pm)$ and $L_{N-3}(3, \lambda_\pm)$. Since the Verma module $M(4,0)$ is irreducible at generic $c$, we have $L(4,0)\equiv M(4,0)$ and its conformal weight $N$ subspace will be $L_{N-4}(4,0)$. The dimensions of these fixed-level subspaces are given by
\begin{equation}
\begin{split}
    d_1 &\coloneqq \dim\, L_{N-1}(1,w_\pm) = P_2(N-1) - 2P_2(N-3) + P_2(N-4)~,\\ 
    d_3 &\coloneqq \dim\, L_{N-3}(3,\lambda_\pm) = P_2(N-3) - P_2(N-4)~,\\
    d_4 &\coloneqq \dim\, L_{N-4}(4,0)=P_2(N-4)~.
\end{split}
\end{equation}
In the following, for reasons of space we will adopt the compact notation,
\begin{equation}
\begin{split}
    \det\, L_{N-1}(1,w_{\pm}) \equiv \det\, L_{N-1}(1,w_{+}) \det\, L_{N-1}(1,w_{-})~,\\
    \det\, L_{N-3}(3,\lambda_{\pm}) \equiv \det\, L_{N-3}(3,\lambda_{+}) \det\, L_{N-3}(3,\lambda_{-})~,
\end{split}
\end{equation} 
\emph{when these determinants appear on the right-hand side} of an equation.

The level $N$ vacuum Kac determinant \eqref{jantzendet} can now be rewritten as: 
\begin{equation}\label{detfull}
\begin{split}
    \det\, V_N(c) &= \lim_{h \to 0}\frac{\det\,G^{(N)}(h,0)}{(h\tau_{1,0})^{2d_1} \det\, L_{N-1}(1,w_{\pm}) (h^2\tau_{3,0})^{2d_3} \det\, L_{N-3}(3,\lambda_{\pm})} \\
    &\qquad\qquad\times \frac{1}{(h^3 \tau_{4,0})^{d_4} \det\, L_{N-4}(4+h, 0)}~.
\end{split}
\end{equation}
Here and below, the factors $\tau_{i,j}$ are ($c$-dependent) normalisation factors given by the Gram determinant of the conformal weight $i$ singular subspace when realised within a Verma module of highest conformal weight $j$ (in a unitriangular basis with respect to a PBW ordering on that Verma module). 

Now one similarly obtains Kac determinants for the irreducible quotients appearing in the denominator in terms of Verma module Kac determinants as follows, 
\begin{equation}
\begin{split}
    \det\, L_{N-1}(1,w_{\pm}) &\sim \lim_{h \to 0} \frac{\det\,G^{(N-1)}(1+h,w_\pm)}{(h\tau_{3,1_\pm})^{2d_3} \det\, L_{N-3}(3,\lambda_{\pm}) (h^2 \tau_{4,1_\pm})^{d_4} \det\, L_{N-4}(4+h, 0)}~,\\ 
    \det\, L_{N-3}(3,\lambda_{\pm}) &\sim \lim_{h \to 0} \frac{\det\,G^{(N-3)}(3+h,\lambda_{\pm})}{(h \tau_{4,3_\pm})^{d_4} \det\, L_{N-4}(4+h, 0)}~,\\
    \det\, L_{N-4}(4+h, 0) &\sim \det\, G^{(N-4)}(4+h,0)~.
\end{split}
\end{equation}
Substituting these into \eqref{detfull} leads to the final answer
\begin{align}
    \det\,(V_N(c)) \sim \lim_{h \to 0} \frac{\mathcal{N}}{h^{2d_1+d_4}} \frac{\det G^{(N)}(h,0)\det G^{(N-3)}(3+h,\lambda_+)^2}{\det G^{(N-1)}(1+h,w_+)^2\det G^{(N-4)}(4+h,0)}~.
\end{align}
Note that $2d_1 + d_4 = \alpha$ in \eqref{vacdet}. $\mathcal{N}$ is the normalisation factor given by: 
\begin{equation}
\begin{split}
    \mathcal{N} &= \frac{(\tau_{3,1_+} \tau_{3,1_-})^{2d_3} (\tau_{4,1_+} \tau_{4,1_-})^{d_4}}{\tau_{1,0}^{2d_1} \tau_{3,0}^{2d_3} \tau_{4,0}^{d_4} (\tau_{4,3_+} \tau_{4,3_-})^{d_4}} \\ &= \frac{1}{(\tau_{1,0}^2)^{P_2(N-1)}} \left( \frac{(\tau_{3,1_+} \tau_{3,1_-})^{2} \tau_{1,0}^4}{\tau_{3,0}^2}\right)^{P_2(N-3)} \left( \frac{\tau_{3,0}^2 \tau_{4,1_+} \tau_{4,1_-}}{\tau_{4,0} \tau_{4,3_+} \tau_{4,3_-} (\tau_{3,1_+} \tau_{3,1_-})^2 \tau_{1,0}^2} \right)^{P_2(N-4)}~.
\end{split}
\end{equation}
Hence this precisely gives the vacuum Kac determinant \eqref{vacdet} up to the determination of the $c$-dependent normalisation factors. The factors $f_1(c), f_3(c)$ and $f_4(c)$ can be fixed in two different ways. Firstly by an explicit computation of $\tau_{i,j}$ factors (by directly identifying the relevant singular subspaces, in an appropriate basis, of the various Verma modules). For example, by such an explicit computation, we find: 
\begin{equation}
\begin{split}
    \tau_{1,0}^2 &= \frac{2-c}{22+5 c}~,\\
    \tau_{3,1_+}^2 &= \tau_{3,1_-}^2 = \frac{(c^2-124c-44)(2-c)}{(22+5c)^3}~,\\
    \tau_{3,0}^2 &= \frac{(98-c)(10+c)^2(2-c)^2}{(22+5c)^5}~.
\end{split} 
\end{equation}
This precisely reproduces the expressions for $f_1(c)$ and $f_3(c)$ given in the main text. Alternatively, a more practical method is to perform explicit low-level computations of the vacuum Kac determinant. In the case of the ${\mathcal W}_3$ algebra, we have three normalisation factors $f_1(c), f_3(c)$ and $f_4(c)$, which can be fixed by comparing \eqref{vacdet} with the explicit computation of the level two, three, and four vacuum Kac determinants, after which the formula \eqref{vacdet} holds for all higher levels. 

\section{\label{app2}Simplifying the vacuum determinant formula}

In this appendix, we describe in detail the simplification of the vacuum module Kac determinant formula \eqref{vacdet} into an explicit rational function of the central charge. To this end, we will need a few observations regarding the Verma module Kac determinant \eqref{Vermadet}. It is immediate to see that the zeros (as a function of $c$) of the leading term in $h$ in the Verma module determinants $\det\ G^{(N)}(h,0), \det\ G^{(N-1)}(1+h,w_\pm), \det\ G^{(N-3)}(3+h,\lambda_\pm)$, and $\det\ G^{(N-4)}(4+h,0)$ fall into four classes: 
\begin{itemize}
    \item $c = c_{p,q}$ with $q > p \geqslant 2$ and $(p,q) = 1$,
    \item $c = c_{1,q}$ with $q \geqslant 2$,
    \item $c = c_{-1,q}$ with $q \geqslant 1$,
    \item $c = c_{p,p} = 2$.
\end{itemize}
By explicit computation, we will see that the vacuum determinant only has zeros that fall into the first case above, eventually leading to the derivation of the simplified expression \eqref{rationalc}. 

The leading term in $h$ in the Verma module Kac determinants $\det\ G^{(N)}(h,0;c)$, $\det\ G^{(N-1)}(1+h,w_\pm)$, $\det\ G^{(N-3)}(3+h,\lambda_\pm)$, and $\det\ G^{(N-4)}(4+h,0)$ are given by
\begin{equation} \label{det0134}
    \begin{split}
        \det G^{(N)}(h,0) &\sim \frac{h^{2 P_2(N-1) + P_2(N-4)}}{(22+5 c)^{s(N)}} \prod_{\substack{q>p\geqslant 1\\(p,q) = 1}} (c - c_{p,q})^{E_{p,q}^{(N),0}} (c_{-1,q}-c)^{P_2(N-(q+2))} (2-c)^{\beta_0}~,\\
        \det G^{(N-1)}(1+h,w_\pm) &\sim \frac{h^{2 P_2(N-3)}}{(22+5 c)^{s(N-1)}} \left( \frac{c^2 - 124 c - 44}{(10+c)^2} \right)^{P_2(N-3)} \\ &\times \prod_{\substack{q>p\geqslant 1\\(p,q) = 1}} (c - c_{p,q})^{E_{p,q}^{(N),1}} \prod_{q \geqslant 2} (c_{-1,q}-c)^{P_2(N-(q+2))} (2-c)^{\beta_1}~, \\ 
        \det G^{(N-3)}(3+h,\lambda_\pm) &\sim \frac{h^{P_2(N-4)}}{(22+5 c)^{s(N-3)}} (c-146)^{P_2(N-4)} \\ &\times \prod_{\substack{q>p\geqslant 1\\(p,q) = 1}} (c - c_{p,q})^{E_{p,q}^{(N),3}} \prod_{q \geqslant 3} (c_{-1,q}-c)^{P_2(N-(q+2))} (2-c)^{\beta_3}~, \\
        \det G^{(N-4)}(4+h,0) &\sim \frac{1}{(22+5 c)^{s(N-4)}} \prod_{\substack{q>p\geqslant 1\\(p,q) = 1}} (c - c_{p,q})^{E_{p,q}^{(N),4}} \\ &\times \prod_{q \geqslant 3} (c_{-1,q}-c)^{P_2(N-(q+2))} (2-c)^{\beta_4}~,
    \end{split}
\end{equation}
where $s(N)$ has been defined in \eqref{sdefn}. The exponents of various $c_{p,q}$ zeros with $q > p \geqslant 1$ and $(p,q) = 1$ appearing in these formulae are given by: 
\begin{equation}
    \begin{split}
        E_{p,q}^{(N),0} &= \sum_{j \in \mathbb{Z}/\{0\}}^\prime (P_2(N - (jp-2)(jq-2)) + 2 P_2(N - (jp-1)(jq-1)))~,\\
        E_{p,q}^{(N),1} &= \sum_{j \in \mathbb{Z}/\{0\}}^\prime (P_2(N-1-(jp-2)(jq-1)) +P_2(N-1-(jp-1)(jq-2)) \\ 
        &+ P_2(N-1-(jp-1)(jq+1)))~, \\ 
        E_{p,q}^{(N),3} &= \sum_{j \in \mathbb{Z}/\{0\}}^\prime (P_2(N-3-(jp-2)(jq+1)) + P_2(N-3-(jp+1)(jq-2)) \\
        &+ P_2(N-3-(jp-1)(jq-1)))~, \\
        E_{p,q}^{(N),4} &= \sum_{j \in \mathbb{Z}/\{0\}}^\prime (P_2(N - 4 - (jp-2)(jq+2)) + 2 P_2(N - 4 - (jp-1)(jq+1)))~,
    \end{split}
\end{equation}
while the exponents of $(2-c)$ are
\begin{equation}
    \begin{split}
        \beta_0 &= P_2(N-1) + 2 P_2(N-4) + 3 \sum_{j \geqslant 3} P_2(N - j^2)~,\\
        \beta_1 &= P_2(N-3) + P_2(N-4) + \sum_{j \geqslant 2} (P_2(N-1-j(j+2)) + 2 P_2(N - 1 - (j(j+1)))~, \\ 
        \beta_3 &= 2\sum_{j \geqslant 1} P_2(N - 3 - j(j+3)) + \sum_{j \geqslant 2} P_2(N - 3 - j^2)~, \\
        \beta_4 &= \sum_{j \geqslant 1}\left( P_2(N - 4 - j(j+4)) + 2 P_2(N - 4 - j(j+2))\right)~.
    \end{split}
\end{equation}
In the sum expressions for the exponents $E_{p,q}^{(N),i}$ there is a restriction (represented by the prime on the summation). For a given $p$, $q$, and $N$, a term in the summand of the form $P_2(N-k-ab)$ where $a$ and $b$ are each one of $jp - 1$, $jq - 1$, $jp - 2$, $jq - 2$, should be considered to appear \emph{only when $ab>0$}.

These can all be determined in a fairly straightforward manner from the explicit expressions of the Verma module Kac determinants upon extracting the leading term in $h$. Since the Verma module $M(4,0;c)$ has no singular vectors at generic $c$, the corresponding Kac determinants $G^{(N-4)}(4+h,0)$ do not vanish in the $h \to 0$ limit. 

Substituting the above expressions into the vacuum determinant formula \eqref{vacdet}, we see that the $h \to 0$ limit is indeed finite and non-zero. Moreover, it is also clear that the exponents of $(c_{-1,q} - c)$ precisely cancel once the normalisation factors are taken into account.\footnote{For $q > 2$, this is obvious. For $q = 1$ and $q = 2$ the cancellation occurs after taking into account the additional $(c_{-1,q}-c)$ factors, namely $(98-c)^{-P_2(N-3)}$ and $(110-c)^{P_2(N-4)}$, arising from the normalisation factors in \eqref{norm}.} It further happens that the powers of $(2-c)$ exactly cancel. To this end, note that
\begin{equation}
    \begin{split}
        &\beta_0 - 2\beta_1 + 2\beta_3 - \beta_4 = P_2(N-1) \textcolor{blue}{+ 2 P_2(N-4) + 3 \sum_{j \geqslant 3} P_2(N - j^2) -  2\sum_{j \geqslant 1} (P_2(N-1-j(j+2))} \\ & \qquad \textcolor{red}{-4 \sum_{j \geqslant 2} P_2(N - 1 - (j(j+1)))} - 2 P_2(N-3) \textcolor{red}{+ 4 \sum_{j \geqslant 1} P_2(N - 3 - j(j+3))}  \\&  \qquad \textcolor{green}{+2 \sum_{j \geqslant 2} P_2(N - 3 - j^2)} \textcolor{blue}{-\sum_{j \geqslant 1}P_2(N - 4 - j(j+4))} \textcolor{green}{-2 \sum_{j \geqslant 1} P_2(N - 4 - j(j+2))} \\& \qquad = P_2(N-1) - 2 P_2(N-3)~,
    \end{split}
\end{equation}
where sub-expressions of a given colour cancel. There is additionally a factor of $(2-c)^{-P_2(N-1)+2P_2(N-3)}$ from  $f_1(c)$ and $f_3(c)$, thus removing all powers of $(2-c)$ in the full vacuum determinant.

A similar analysis of the exponent of $(c - c_{1,q})$ reveals an analogous cancellation of these terms for all $q \geqslant 2$. In particular, manipulations of two-coloured partitions show that $E_{1,q}^{(N),0} - 2 E_{1,q}^{(N),1} + 2 E_{1,q}^{(N),3} - E_{1,q}^{(N),4} = 0$ for all $q > 2$. The case of $q = 2$ is slightly subtle. In this case, one evaluates the exponent of $(c+10)$ to be $E_{1,q}^{(N),0} - 2 E_{1,q}^{(N),1} + 2 E_{1,q}^{(N),3} - E_{1,q}^{(N),4} + 4 P_2(N-3) =  2 P_2(N-3) + P_2(N-4)$, which again precisely cancels the factor of $(c+10)^{-2P_2(N-3) - P_2(N-4)}$ arising from $f_3(c)$ and $f_4(c)$ in \eqref{norm}. The conclusion is that the level $N$ vacuum determinant only has zeros at $c = c_{p,q}$ with $q > p \geqslant 2$ and $(p,q) = 1$, and the multiplicity of these zeros is given by:
\begin{equation}
    E_{p,q}^{(N)} \coloneqq E_{p,q}^{(N),0} - 2 E_{p,q}^{(N),1} + 2 E_{p,q}^{(N),3} - E_{p,q}^{(N),4}
\end{equation}
along with the additional $c$ dependent term of $(22+5c)^{n(N)}$, thus reducing the expression \eqref{vacdet} to the rational expression \eqref{rationalc}. 

\bibliographystyle{aux/JHEP}
\bibliography{aux/references}

\end{document}